\documentclass{amsart}
\def\Bshorttitle{FLUID LIMIT OF THE ENSKOG EQUATION WITH GUARANTEED H-THEOREM}
\def\Bauthor{AOTO TAKAHASHI AND SHIGERU TAKATA}
\usepackage{hyperref}
\usepackage{amsmath}
\usepackage{amssymb}
\usepackage{amsfonts}
\usepackage{bm}
\usepackage{amsthm}
\usepackage{array,epsfig,fancyheadings}%,showlabels}
\usepackage[sectionbib,numbers]{natbib}
\skip\footins =.6cm
\begin{document}
\allowdisplaybreaks \pagestyle{fancy} \thispagestyle{fancyplain}
\lhead[\fancyplain{}\leftmark]{} \chead[]{}
\rhead[]{\fancyplain{}\rightmark} \cfoot{\rm\thepage}
%\headrulewidth=0pt %New Addition
\def\n{\noindent}
\newtheorem{theorem}{Theorem}
\newtheorem{lemma}{Lemma}
\newtheorem{corollary}{Corollary}
\newtheorem{proposition}{Proposition}
\theoremstyle{definition}
\newtheorem{definition}{Definition}
\newtheorem{example}{Example}
\newtheorem{remark}{Remark}
\newcount\fs
\def\dd{\kern 2pt\raise 3pt\hbox{\large .}\kern 2pt\ignorespaces}
\def\fontsizefi{\fontsize{5}{9.5pt plus.8pt minus .6pt}\selectfont}
\def\fontsizesi{\fontsize{6}{10.5pt plus.8pt minus .6pt}\selectfont}
\def\fontsizese{\fontsize{7}{10.5pt plus.8pt minus .6pt}\selectfont}
\def\fontsizeei{\fontsize{8}{12pt plus.8pt minus .6pt}\selectfont}
\def\fontsizeni{\fontsize{9}{13.5pt plus1pt minus .8pt}\selectfont}
\def\fontsizete{\fontsize{10}{14pt plus.8pt minus .6pt}\selectfont}
\def\fontsizeel{\fontsize{10.95}{15pt plus2pt minus .0pt}\selectfont}
\def\fontsizetw{\fontsize{12}{116.5pt plus1pt minus .8pt}\selectfont}
\def\fontsizefo{\fontsize{14.4}{18pt plus1pt minus .8pt}\selectfont}
\def\fontsizesn{\fontsize{17.28}{20pt plus1pt minus 1pt}\selectfont}
\def\fontsizetn{\fontsize{20.73}{26pt plus1pt minus 1pt}\selectfont}
\def\fontsizetf{\fontsize{24.88}{33pt plus1.5pt minus 1pt}\selectfont}
\def\sz#1#2{\fs=#1#2
   \ifnum\fs=05 \fontsizefi
   \else\ifnum\fs=06 \fontsizesi
   \else\ifnum\fs=07 \fontsizese
   \else\ifnum\fs=08 \fontsizeei
   \else\ifnum\fs=09 \fontsizeni
   \else\ifnum\fs=10 \fontsizete
   \else\ifnum\fs=11 \fontsizeel
   \else\ifnum\fs=12 \fontsizetw
   \else\ifnum\fs=14 \fontsizefo
   \else\ifnum\fs=17 \fontsizesn
   \else\ifnum\fs=20 \fontsizetn
   \else\ifnum\fs=25 \fontsizetf
\fi\fi \fi \fi \fi \fi \fi \fi \fi \fi \fi \fi }

%	%	%	%	%
\newcommand{\hC}{{\hat{c}}}
\newcommand{\hbC}{{\hat{\bm{c}}}}
\newcommand{\he}{{\hat{e}}}
\newcommand{\hp}{{\hat{p}}}
\newcommand{\hr}{{\hat{\rho}}}
\newcommand{\hmu}{{\hat{\mu}}}
\newcommand{\hmR}{{\hat{\mathcal{R}}}}
\newcommand{\hmS}{{\hat{\mathcal{S}}}}
\newcommand{\hq}{{\hat{q}}}
\newcommand{\hT}{{\hat{T}}}
\newcommand{\hD}{{\hat{D}}}
\newcommand{\hf}{{\hat{f}}}
\newcommand{\hg}{{\hat{g}}}
\newcommand{\hh}{{\hat{h}}}
\newcommand{\hlam}{{\hat{\lambda}}}
\newcommand{\hsg}{{\hat{\sigma}}}
\newcommand{\whsg}{{\widehat{\sigma}}}
\newcommand{\hJ}{{\hat{J}}}
\newcommand{\htt}{{\hat{t}}}
\newcommand{\hv}{{\hat{v}}}
\newcommand{\hU}{{\hat{U}}}
\newcommand{\hbU}{{\hat{\bm{U}}}}
\newcommand{\hbV}{{\hat{\bm{V}}}}
\newcommand{\hbW}{{\hat{\bm{W}}}}
\newcommand{\hV}{{\hat{V}}}
\newcommand{\hW}{{\hat{W}}}
\newcommand{\bx}{{\bm{x}}}
\newcommand{\by}{{\bm{y}}}
\newcommand{\bX}{{\bm{X}}}
\newcommand{\bY}{{\bm{Y}}}
\newcommand{\bxi}{{\bm{\xi}}}
\newcommand{\bz}{{\bm{\zeta}}}
\newcommand{\hbv}{{\hat{\bm{v}}}}
\newcommand{\bu}{{\bm{u}}}
\newcommand{\bv}{{\bm{v}}}
\newcommand{\bhu}{{\hat{\bm{u}}}}
%\newcpmmand{\ds}{\displaystyle}
\newcommand{\hspa}{{\hat{s}}}
\newcommand{\spa}{{s}}

\newcommand{\sfg}{{\mathsf{g}}}
\newcommand{\mF}{{\mathcal{F}}}
\newcommand{\mG}{{\mathcal{G}}}
\newcommand{\mC}{{\mathcal{C}}}
\newcommand{\mJ}{{\mathcal{J}}}
\newcommand{\mL}{{\mathcal{L}}}
\newcommand{\mR}{{\mathcal{R}}}
\newcommand{\mS}{{\mathcal{S}}}
\newcommand{\mY}{{\mathcal{Y}}}
\newcommand{\Kn}{{\mathrm{Kn}}}
\newcommand{\hmY}{{\hat{\mathcal{Y}}}}

\newcommand{\kin}{{\mathrm{kin}}}
\newcommand{\col}{{\mathrm{col}}}
\newcommand{\ve}{{\varepsilon}}
\newcommand{\vr}{{\varrho}}
\newcommand{\nbvr}{{\nabla\varrho}}
\newcommand{\z}{{\zeta}}

%	%	%	%	%

\markright{{\scriptsize $\ $}\hfill\break\vskip -16pt {\scriptsize
$\ $}\hfill\break\vskip -16pt {\scriptsize $\ $} \hfill}

\markboth {\hfill{\small \rm \Bauthor}\hfill} {\hfill {\small \rm
\Bshorttitle} \hfill}
\renewcommand{\thefootnote}{}
\abovedisplayskip 12pt
\belowdisplayskip 12pt
\abovedisplayshortskip 5pt
\belowdisplayshortskip 9pt

$\ $\par \vskip .5cm \centerline{\large\bf Fluid-dynamic limit of the Enskog equation} \smallskip
 \centerline{\large\bf  with the guaranteed H-theorem}\vskip .5cm

\centerline{AOTO TAKAHASHI$^{1,a}$ AND SHIGERU TAKATA$^{1,b}$}
\vskip .5cm
\emph{Dedicated to Professor Shih-Hsien Yu on the occasion of his 60th birthday}
\vskip .5cm
\n $^1$Department of Aeronautics and Astronautics, Graduate School of Engineering, Kyoto University, Kyoto 615-8540, Japan   \vskip -4pt
 \noindent
$^a$E-mail: \href{mailto:takahashi.aoto.63c@st.kyoto-u.ac.jp}{takahashi.aoto.63c@st.kyoto-u.ac.jp} \vskip -4pt
\n 
%$^2$Kyoto University \vskip -4pt
 \noindent
$^b$E-mail: \href{mailto:takata.shigeru.4a@kyoto-u.ac.jp}{takata.shigeru.4a@kyoto-u.ac.jp} (Corresponding author)

\footnote{\footnotesize\hskip -.5cm AMS
Subject Classification: 82C40, 76P99.} \footnote{\footnotesize\hskip -.5cm
Key words and phrases: Enskog equation, dense gas, fluid-dynamic limit, Chapman--Enskog method}

\centerline{\bf Abstract}\par \vskip .2cm 
The fluid-dynamic limit of the Enskog equation with a slight modification is discussed on the basis of the Chapman--Enskog method. 
This modified version of the Enskog equation has been shown recently by the present authors to ensure the H-theorem.
In the present paper, it is shown that the modified version recovers the same fluid-dynamic description of the dense gas as the original Enskog equation, at least up to the level of the Navier--Stokes--Fourier set of equations inclusive. 
Since the original Enskog equation is known to recover the fluid-dynamical transport properties well, this result implies that the modified version of the Enskog equation provides consistent descriptions both thermodynamically and fluid-dynamically.
 \par

\sz11 \section{Introduction\label{sec:intro}}

It is widely known that 
the Boltzmann equation well describes the behavior of dilute gases. 
Such gases behave as ideal gases and the corresponding fluid-dynamic limit,
i.e., the small mean-free-path (or the small Knudsen number) limit,
of the Boltzmann equation has been intensively studied 
in physics, applied mathematics, and engineering communities.
However, the Boltzmann equation ceases its applicability to dense gases
in which the so-called non-ideal gas effect manifests itself.
In such circumstances, the details of collision dynamics and collision probability based on the one-particle distribution function have to be reconsidered.
Enskog derived a kinetic equation for dense gases \cite{E72}
by modifying the Boltzmann equation mainly in two points:
(i) the centers of two colliding molecules are separated by the distance of the molecular diameter;
(ii) there is a correlation between the one-particle velocity distribution functions (VDFs) of two colliding particles. 
The kinetic equation taking care of these points is nowadays called the Enskog equation. Enskog also discussed the transport properties of dense gases 
in terms of the fluid-dynamic limit of his equation on the basis of the Chapman--Enskog method.

Although Enskog successfully derived the kinetic equation for dense gases,
the correlation mentioned in (ii) was determined somehow intuitively and semi-phenomenologically.
Indeed, Enskog chose the correlation factor in (ii) as a function of the number (or mass) density at the midpoint of the colliding pair of molecules.
In the semi-phenomenological view, the form of the function is adjustable to recover the desired equation of state (EoS) for the dense gas. 
However, the H-theorem has never been established for the original form of his equation.

van Beijeren \& Ernst \cite{BE73} proposed a modification of the correlation factor that takes account of the many-body configuration of hard spheres and named it the modified (or revised) Enskog equation (MEE).  It was later shown by Resibois \cite{R78} to ensure the H-theorem in a periodic domain. It is this point, in the case of a single-component gas, that MEE has an advantage over the original Enskog equation (OEE), although the MEE was originally motivated to overcome the inconsistency of OEE related to the Onsager reciprocity.
The MEE is thus theoretically satisfactory,
but the correlation factor in the MEE has a series structure of rapidly complicated and has been hindering its further applications. 
It is not easy even to find the EoS for the gas described by the MEE. 
Rather recently, Benilov \& Benilov \cite{BB18} modified the series in MEE from the semi-phenomenological perspective so that it can be adjusted to the desired EoS, while ensuring that the H-theorem holds. 
Nevertheless, the rapidly complicated structure of the series is retained,
which still hinders straightforward practical applications. 
The many-body configuration feature embedded in the correlation factor
undermines the simplicity of the description in terms of the one-particle VDF.

We have recently found \cite{TT25} that
the H-theorem can be established by only a slight modification of Enskog's original
correlation factor. The novel modification does not rely on the series structure 
and is adjustable to the desired EoS.
In the present paper, we will present this Enskog equation with slight modification (EESM) 
and discuss its fluid-dynamic limit in details, as announced in \cite{TT25}, on the basis of the Chapman--Enskog method following the procedure by Grad \cite{G58,S07}.
%This is part of our project on dense gas kinetics toward advancing the description of the phase transition responsible for evaporation and condensation gas flows. We are pleased to dedicate the present work to Prof. Shih-Hsien Yu on the occasion of his 60th birth day, who developed in collaboration with Tai-Ping Liu the theory of invariant manifolds for the Boltzmann equation that led to a mathematical understanding of the striking bifurcation in the transition of evaporation to condensation.

\section{Enskog equation and equation of state: generic description}

We consider the Enskog equation for a single species dense gas
that is composed of hard-sphere molecules with a common diameter $\sigma$ and mass $m$
in a spatial domain $D$, where the centers of the molecules can be located. 
Let $t$, $\bm{X}$, and $\bm{\xi}$
be a time, a spatial position, and a molecular velocity, respectively.
Denoting the one-particle distribution function of gas molecules
by $f(t,\bm{X},\bm{\xi}$), 
the Enskog equation is written
as 
\begin{subequations}\label{eq:MEE}
\begin{align}
 & \frac{\partial f}{\partial t}+\xi_{i}\frac{\partial f}{\partial X_{i}}=J(f)\equiv J^{G}(f)-J^{L}(f),\quad \mathrm{for\ }\bm{X}\in D,\displaybreak[0]\label{eq:2.1}\\
 & J^{G}(f)\equiv\frac{\sigma^{2}}{m}\int {g(\bm{X}_{\sigma\bm{\alpha}}^{+},\bm{X})f_{*}^{\prime}(\bm{X}_{\sigma\bm{\alpha}}^{+})f^{\prime}(\bm{X})}V_{\alpha}\theta(V_{\alpha})d\Omega(\bm{\alpha})d\bm{\xi}_{*},\displaybreak[0]\label{eq:2.2}\\
 & J^{L}(f)\equiv\frac{\sigma^{2}}{m}\int {g(\bm{X}_{\sigma\bm{\alpha}}^{-},\bm{X})f_{*}(\bm{X}_{\sigma\bm{\alpha}}^{-})f(\bm{X})}V_{\alpha}\theta(V_{\alpha})d\Omega(\bm{\alpha})d\bm{\xi}_{*},\label{eq:2.3}
\end{align}
\end{subequations}
\noindent
where $\bm{X}_{\bm{r}}^{\pm}=\bm{X}\pm\bm{r}$,
$\bm{\alpha}$ is a unit vector, 
$d\Omega(\bm{\alpha})$ is a
solid angle element in the direction of $\bm{\alpha}$,
$\theta$ is the Heaviside function
\begin{subequations}
\begin{equation}
\theta(x)=\begin{cases}
1, & x\ge0\\
0, & x<0
\end{cases},
\end{equation}
\noindent
and the following notation convention has been used:
\begin{align}
 & \begin{cases}
{
f(\bm{X})=f(\bm{X},\bm{\xi}),\ f^{\prime}(\bm{X})=f(\bm{X},\bm{\xi}^{\prime})},\\
{
f_{*}(\bm{X}_{\sigma\bm{\alpha}}^{-})=f(\bm{X}_{\sigma\bm{\alpha}}^{-},\bm{\xi}_{*}),\ f_{*}^{\prime}(\bm{X}_{\sigma\bm{\alpha}}^{+})=f(\bm{X}_{\sigma\bm{\alpha}}^{+},\bm{\xi}_{*}^{\prime})},
\end{cases}\label{eq:contf}\displaybreak[0]\\
 & \bm{\xi}^{\prime}=\bm{\xi}+V_{\alpha}\bm{\alpha},\quad\bm{\xi}_{*}^{\prime}=\bm{\xi}_{*}-V_{\alpha}\bm{\alpha},\quad V_{\alpha}=\bm{V}\cdot\bm{\alpha},\quad\bm{V}=\bm{\xi_{*}}-\bm{\xi}.\label{eq:2.5}
\end{align}
\noindent
Here and in what follows, the argument $t$ is often suppressed, unless confusion is anticipated. 
The convention \eqref{eq:contf} will be applied only to the quantities that depend on molecular velocity.
It should be noted that \eqref{eq:MEE} makes sense only 
when the positions $\bm{X}$, $\bm{X}^+_{\sigma\bm{\alpha}}$,
and $\bm{X}^-_{\sigma\bm{\alpha}}$ are all in the domain $D$,
which may restrict the range of integration 
with respect to $\bm{\alpha}$ and $\bm{\xi}_*$.
However, by including the indicator function $\chi_D$, 
\begin{equation}
\chi_{D}(\bm{X})=\begin{cases}
1, & \bm{X}\in D\\
0, & \mbox{otherwise}
\end{cases},\label{eq:chi_def}
\end{equation}
\noindent
in the definition of $g$ in such a way that
\begin{equation}
g(\bX,\bY)=\sfg(\bX,\bY)\chi_D(\bX)\chi_D(\bY),\label{eq:sfg}
\end{equation}%
\end{subequations}
\noindent
the range of integration in
(\ref{eq:2.2}) and (\ref{eq:2.3}) can be treated as the whole space
of $\bm{\xi}_*$ and all directions of $\bm{\alpha}$, irrespective of the position in the domain $D$.

The factor $\sfg$ occurring in \eqref{eq:sfg} is generically
assumed to be positive and symmetric with respect to the exchange of two position vectors: $\sfg(\bm{X},\bm{Y})=\sfg(\bm{Y},\bm{X})$.
There are some varieties of $\sfg$ in the literature.
However, except for the forms based on the many-body configurations such as
\begin{subequations}
\begin{multline}
\sfg(\bm{r}_1,\bm{r}_2)
=1+\int \frac{\rho(\bm{r}_3)}{m}\mathcal{V}(\bm{r}_1\bm{r}_2|\bm{r}_3)d\bm{r}_3 \\
  +\frac12\int \frac{\rho(\bm{r}_3)}{m}\frac{\rho(\bm{r}_4)}{m}
   \mathcal{V}(\bm{r}_1\bm{r}_2|\bm{r}_3\bm{r}_4)d\bm{r}_3d\bm{r}_4
+\cdots,\label{eq:g_MEE}
\end{multline}
\noindent
in \cite{BE73,DVK21} 
and
\begin{multline}
\sfg(\bm{r},\bm{r}_1)=
1+\sum_{l=1}^\infty c_l\int
\Big[ \prod_{i=2}^{l+1}\prod_{j=i+1}^{l+1}
      \theta(\sigma-|\bm{r}_i-\bm{r}_j|)
\Big]\\
\times
\Big[ \prod_{i=2}^{l+1} \frac{\rho(\bm{r}_i)}{m}
      \theta(\sigma-|\bm{r}-\bm{r}_i|)\theta(\sigma-|\bm{r}_1-\bm{r}_i|)
\Big]\prod_{i=2}^{l+1}d\bm{r}_i,\label{eq:g_BB}
\end{multline}
\end{subequations}
\noindent
in \cite{BB18,BB19},
the H-theorem has not been established for a long time for the Enskog equation.
In the above,  
$\mathcal{V}(\bm{r}_1\bm{r}_2|\bm{r}_3\cdots\bm{r}_l)$ is the sum of all Mayer graphs of $l$ labelled points \cite{BE73,DVK21},
$c_l$ is a constant adjustable to a given EoS,
and $\rho$ is the mass density defined by
\begin{equation}
\rho(t,\bm{X})=\int f(t,\bm{X},\bm{\xi})d\bm{\xi}.
\label{eq:rho_def}
\end{equation}

In contrast to the complexity of the above successful forms in \cite{BE73,BB18},
we have quite recently found \cite{TT25}
that the H-theorem can be established for the Enskog equation 
with the following much simpler form of $\sfg$:
\begin{subequations}\begin{align}
& \sfg(\bm{X},\bm{Y})={\mS}({\mR}(\bm{X}))+{\mS}({\mR}(\bm{Y})),
\label{eq:g_def}
\\
& {\mR}(\bm{X})=\frac{1}{m}\int_D \rho(\bm{Y})\theta(\sigma-|\bm{Y}-\bm{X}|) d\bm{Y},
\label{eq:R_def}
\end{align}\end{subequations}
\noindent
where $\mS$ is non-negative.
Since $\mR$ is dimensionless, $\mS$ as well as $\sfg$ are also dimensionless.
The specific form of ${\mS}$ can be determined
in accordance with the EoS of the gas from the semi-phenomenological perspective,
as in the case of the original Enskog equation.
%In addition to the simple form,
%the straightforward connection to the EoS 
%is the benefit of this novel form over the previous ones
%that relies on the series structure coming from many-body configuration.
In the original Enskog equation,
the factor $\sfg$ was given as
\begin{equation}
\sfg(\bX,\bY)=\mY(\rho(\frac{\bX+\bY}{2})),\label{eq:gY}
\end{equation}
\noindent
and the specific form of $\mY$ can be chosen semi-phenomenologically
in accordance with the EoS under consideration as well. 
Thus, the novel $\mS$, like the conventional $\mY$, has the advantage
of a simple connection to a given EoS,
compared with \eqref{eq:g_MEE} and \eqref{eq:g_BB}.

In closing this section, we list the definitions
of macroscopic quantities for later convenience.
In addition to the density $\rho$ already given in \eqref{eq:rho_def},
the flow velocity $\bm{v}$ (or $v_i$) and temperature $T$
are defined by
\begin{subequations}
\begin{equation}
 v_i=\frac{1}{\rho}\int \xi_i f d\bxi, \quad
 T=\frac{1}{3R\rho}\int (\bm{\xi}-\bm{v})^2 f d\bxi, 
\end{equation}
\noindent
with $R$ being the specific gas constant
(the Boltzmann constant $k_B$ divided by $m$), while the specific internal energy $e$,
the so-called kinetic part of the stress tensor $p_{ij}^{\kin}$, and that of the heat-flow vector $\bm{q}^{\kin}$ (or $q_i^{\kin}$) are defined by
\begin{align}
& e=\frac{1}{2\rho}\int (\bm{\xi}-\bm{v})^2 f d\bxi
  (=\frac32RT), \label{eq:kin_inteng_def}\\
& p_{ij}^{\kin}=\int (\xi_i-v_i)(\xi_j-v_j)f d\bxi, \label{eq:kin_stress_def}\\
& q_i^{\kin}=\frac12\int (\xi_i-v_i)(\bm{\xi}-\bm{v})^2 f d\bxi. \label{eq:kin_heatfl_def}
\end{align}
\end{subequations}
\noindent
Here, the stress tensor and the heat-flow vector are discriminated 
from the contributions from the collision term
that will be explained later in Sec.~\ref{sec:pq_col}.
The internal energy has no contribution from the collision term,
so that the discrimination is not necessary. 

\section{Brief exposition of the H-theorem\label{sec:main}}
\subsection{Kinetic part of the H function}

First we shall focus on the so-called kinetic part of the H function$^1$%
\footnote{$^1$ To be precise, it is necessary to make the argument of the logarithmic function dimensionless, like $\ln (f/c_0)$ with a constant $c_0$ having the same dimension as $f$.
We, however, leave the argument dimensional 
to avoid additional calculations that do not affect the results.\label{fn1}
}
that is defined by
\begin{equation}
\mathcal{H}^{\kin}\equiv\int_{D}\int f{\ln f} d\bxi d\bm{X}.\label{H_kinetic}
\end{equation}
\noindent
The equation for this part can be derived by the standard manipulation
for the Boltzmann equation as
\begin{equation}
\frac{d}{d t}\mathcal{H}^{\kin}
+\int_D\frac{\partial H_i^\kin}{\partial X_{i}} d\bm{X}
=\int_{D} \int J(f){\ln f} d\bxi d\bm{X},\label{eq:H^k}
\end{equation}
\noindent
with
\begin{equation}
H_i^{\kin}=\int \xi_i f\ln f d\bxi,
\end{equation}
\noindent
where the spatial integration over the domain $D$
is necessary for further consideration of the collision term,
which is a marked difference from the case of the Boltzmann equation,
e.g., \cite{R78,BLPT91,MGB18,T24}.
\noindent
Note that, 
thanks to the indicator function $\chi_D$ occurring in $g$ [see \eqref{eq:sfg}], 
the range of integration on the right-hand side
can be changed from $D$ to $\mathbb{R}^3$.
This change of integration range allows the shift and other operations 
 necessary to handle the right-hand side
 (see, e.g., Sec.~4 of \cite{T24} or Appendix~A of \cite{TT25}).
 We omit the details of manipulations
 and only present the final results, i.e.,
\begin{multline}
  \int_{D}\int J(f){\ln f} d\bxi d\bm{X} %\nonumber \\
%= & \int\ln[f^{\prime}(\bm{X}_{\sigma\bm{\alpha}}^{+})/f(\bm{X}_{\sigma\bm{\alpha}}^{+})]g(\bm{X},\bm{X}_{\sigma\bm{\alpha}}^{+})f_{*}(\bm{X})f(\bm{X}_{\sigma\bm{\alpha}}^{+})V_{\alpha}\theta(V_{\alpha})d\Omega(\bm{\alpha})d\bm{\xi}_{*}d\bm{\xi}d\bm{X}\displaybreak[0]\nonumber \\
%= & \int\ln[f_{*}^{\prime}(\bm{X}_{\sigma\bm{\alpha}}^{-})/f_{*}(\bm{X}_{\sigma\bm{\alpha}}^{-})]g(\bm{X},\bm{X}_{\sigma\bm{\alpha}}^{-})f(\bm{X})f_{*}(\bm{X}_{\sigma\bm{\alpha}}^{-})V_{\alpha}\theta(V_{\alpha})d\Omega(\bm{\alpha})d\bm{\xi}d\bm{\xi}_{*}d\bm{X}\displaybreak[0]\nonumber \\
= \frac{\sigma^{2}}{2m}
\int f(\bm{X})f_{*}(\bm{X}_{\sigma\bm{\alpha}}^{-})
\ln\Big(\frac{f_{*}^{\prime}(\bm{X}_{\sigma\bm{\alpha}}^{-})f^{\prime}(\bm{X})}{f_{*}(\bm{X}_{\sigma\bm{\alpha}}^{-})f(\bm{X})}\Big) \\
\times g(\bm{X},\bm{X}_{\sigma\bm{\alpha}}^{-})V_{\alpha}\theta(V_{\alpha})d\Omega(\bm{\alpha})d\bm{\xi}d\bm{\xi}_{*}d\bm{X}.
\label{eq:Jlnf}
\end{multline}
\noindent
Since $x\ln(y/x)\le y-x$ for any $x,y>0$ and the equality holds if and only if $x=y$, we have the estimate \cite{BLPT91} that
\begin{equation}
\int_{D}\int J(f){\ln f} d\bxi d\bm{X}\le I(t),\label{eq:J2I}
\end{equation}
\noindent
where 
\begin{multline}
I(t)=\frac{\sigma^{2}}{2m}\int g(\bm{X},\bm{X}_{\sigma\bm{\alpha}}^{-})[f_{*}^{\prime}(\bm{X}_{\sigma\bm{\alpha}}^{-})f^{\prime}(\bm{X}) \\
-f(\bm{X})f_{*}(\bm{X}_{\sigma\bm{\alpha}}^{-})]V_{\alpha}\theta(V_{\alpha})d\Omega(\bm{\alpha})d\bm{\xi}d\bm{\xi}_{*}d\bm{X},\label{eq:I_def}
\end{multline}
\noindent
and the equality holds if and only if $I(t)=0$ or equivalently
\begin{equation}
f_{*}^{\prime}(\bm{X}_{\sigma\bm{\alpha}}^{-})f^{\prime}(\bm{X})-f(\bm{X})f_{*}(\bm{X}_{\sigma\bm{\alpha}}^{-})=0.
\label{eq:equilibrium}
\end{equation}
\noindent
It should be remarked that, 
as shown in Appendix~B.1 of \cite{TT25},
$I(t)$ is eventually reduced to
\begin{equation}
I(t)=-\frac{\sigma^{2}}{m}\int g(\bm{X},\bm{X}_{\sigma\bm{\alpha}}^+)\rho(\bm{X})\rho(\bm{X}_{\sigma\bm{\alpha}}^{+})
\bm{v}(\bm{X})\cdot\bm{\alpha}d\Omega(\bm{\alpha})d\bm{X}.
\label{eq:I_final}
\end{equation}
\noindent
However, it is not clear whether or not $I(t)$ is non-positive.

\subsection{Collisional part of the H function}

In order to handle $I(t)$,
we newly introduce the following function
\begin{equation}
\mathcal{H}^{\col}(t)=\int_D \rho(\bm{X}) [\int_0^{{\mR}(\bm{X})}
{\mS}(x)dx ] d\bm{X}.\label{eq:H^c}
\end{equation}
\noindent
Then, 
it has been shown in \cite{TT25} that
\begin{equation}
  \frac{d}{dt}\mathcal{H}^{\col}
+\int_D \frac{\partial H_i^\col}{\partial X_i}d\bm{X}=-I(t),\label{eq:H_col}
\end{equation}
\noindent
holds, where
\begin{multline}
H_i^{\col}=\rho(\bm{X}){v}_i(\bm{X})[\int_0^{{\mR}(\bm{X})}{\mS}(x)dx \\
  +\int_D\frac{\rho(\bm{Y})}{m}\theta(\sigma-|\bm{Y}-\bm{X}|){\mS}({\mR}(\bm{Y}))d\bm{Y}].
\end{multline}
\subsection{H-theorem}

From \eqref{eq:H^k} with \eqref{eq:J2I} and \eqref{eq:H_col}, 
the time derivative of the sum
$\mathcal{H}\equiv\mathcal{H}^{\kin}+\mathcal{H}^{\col}$
satisfies that
\begin{equation}
\frac{d\mathcal{H}}{dt}
=\frac{d}{dt}(\mathcal{H}^{\kin}+\mathcal{H}^{\col})
\le 
\int_{\partial D} (H_i^{\kin}+H_i^{\col})n_i dS,
\label{eq:H_inequality}
\end{equation}
\noindent
where $\bm{n}$ (or $n_i$) is the inward unit normal to the boundary 
and $dS$ is the surface element of the boundary $\partial D$.
Note that the equality holds if and only if \eqref{eq:equilibrium} is satisfied.
This condition restricts the VDF to the Maxwellian,
the detailed discussions of which can be found in \cite{TT25}.

The inequality \eqref{eq:H_inequality} 
is the H-theorem for the EESM.
In \cite{TT25}, the monotonicity of $\mathcal{H}$
has been discussed in details for three typical cases: 
the domain $D$ is three dimensional and 
is (i) periodic, (ii) surrounded by the specular reflection boundary,
and (iii) surrounded by the impermeable surface of a heat bath
with a uniform constant temperature $T_w$. 
The aim of the present paper is to investigate the fluid-dynamic behavior
described by EESM, which is thermodynamically consistent.

\section{Collisional contributions to transport properties
and static pressure\label{sec:collision}}

\subsection{Collisional part of the stress tensor and heat-flow vector\label{sec:pq_col}}

One of the relevant consequences different from the Boltzmann equation
is the occurrence of the momentum and the energy transfer
from the collision term of the Enskog equation.
Indeed, the usual symmetry relation for the Boltzmann collision integral
does not hold, but instead only its incomplete form does hold
\begin{multline}
 \int \varphi J(f) d\bxi
=\frac{1}{2}\frac{\sigma^{2}}{m}\int(\varphi-\varphi^{\prime})\{g(\bm{X}_{\sigma\bm{\alpha}}^{+},\bm{X})f(\bm{X}_{\sigma\bm{\alpha}}^{+})f_{*}(\bm{X})\\
-g(\bm{X}_{\sigma\bm{\alpha}}^{-},\bm{X})f_{*}(\bm{X}_{\sigma\bm{\alpha}}^{-})f(\bm{X})\}V_{\alpha}\theta(V_{\alpha})d\Omega(\bm{\alpha})d\bm{\xi}_{*}d\bm{\xi}, \label{eq:varphi_transform}
\end{multline}
\noindent
where $\varphi$ represents an arbitrary function.
As the result, $\int \varphi J(f) d\bm{\xi}$ does not necessarily vanish
even when $\varphi=\xi_i$ and $\bm{\xi}^2$.
This means that the momentum and the energy transport
by means of molecular collisions locally occur.

Furthermore,
the following transformation \cite{CL88,MGB18}
\begin{align}
 & g(\bm{X}_{\sigma\bm{\alpha}}^{-},\bm{X})f_{*}(\bm{X}_{\sigma\bm{\alpha}}^{-})f(\bm{X})-g(\bm{X}_{\sigma\bm{\alpha}}^{+},\bm{X})f(\bm{X}_{\sigma\bm{\alpha}}^{+})f_{*}(\bm{X})\displaybreak[0]\notag\\
= & -\int_{0}^{\sigma}\frac{\partial}{\partial\spa}[g(\bm{X}_{\spa\bm{\alpha}}^{+},\bm{X}_{(\spa-\sigma)\bm{\alpha}}^{+})f_{*}(\bm{X}_{(\spa-\sigma)\bm{\alpha}}^{+})f(\bm{X}_{\spa\bm{\alpha}}^{+})]d\spa\displaybreak[0]\notag\\
= & -\frac{\partial}{\partial X_i}\int_{0}^{\sigma}\alpha_i
 g(\bm{X}_{\spa\bm{\alpha}}^{+},\bm{X}_{(\spa-\sigma)\bm{\alpha}}^{+})f_{*}(\bm{X}_{(\spa-\sigma)\bm{\alpha}}^{+})f(\bm{X}_{\spa\bm{\alpha}}^{+})d\spa,\label{eq:15}
\end{align}
\noindent
leads to alternative expressions of $\int \varphi J(f) d\bm{\xi}$
for $\varphi=\xi_i$ and $\bm{\xi}^2$ \cite{CL88,MGB18}
such that
\begin{subequations}
\begin{align}
 \int \xi_j J(f) d\bxi
= &-\frac{\partial}{\partial X_i}p_{ij}^{\col}, \\
  \frac{1}{2}\int \bm{\xi}^{2}J(f) d\bxi
=&-\frac{\partial}{\partial X_{i}}(p_{ij}^{\col}v_{j}+q_{i}^{\col}),\label{eq:4.8}
\end{align}
\noindent
with
\begin{multline}
p_{ij}^{\col}= 
 \frac{\sigma^{2}}{2m}\int {\int_{0}^{\sigma}}\alpha_{i}\alpha_{j}V_{\alpha}^{2}
 \theta(V_{\alpha}) \\
 \times g(\bm{X}_{\spa\bm{\alpha}}^{+},\bm{X}_{(\spa-\sigma)
 \bm{\alpha}}^{+})f_{*}(\bm{X}_{(\spa-\sigma)\bm{\alpha}}^{+})f(\bm{X}_{\spa\bm{\alpha}}^{+}){d\spa} d\Omega(\bm{\alpha})d\bm{\xi}_{*}d\bm{\xi}, \label{eq:col_stress}
\end{multline}
\begin{multline}
q_{i}^{\col}=\frac{\sigma^{2}}{4m}\int{\int_{0}^{\sigma}}\alpha_{i}[(\bm{c}+\bm{c}_{*})\cdot\bm{\alpha}]V_{\alpha}^{2}\theta(V_{\alpha})
 \\
\times g(\bm{X}_{\spa\bm{\alpha}}^{+},\bm{X}_{(\spa-\sigma)\bm{\alpha}}^{+})f_{*}(\bm{X}_{(\spa-\sigma)\bm{\alpha}}^{+})f(\bm{X}_{\spa\bm{\alpha}}^{+}){d\spa} d\Omega(\bm{\alpha})d\bm{\xi}_{*}d\bm{\xi}.\label{eq:4.6}
\end{multline}\end{subequations}
\noindent
Here,
\begin{equation}
\varphi-\varphi^{\prime}
=\begin{cases}
-V_{\alpha}\alpha_{i}, & {\displaystyle (\varphi=\xi_{i}),}\\
{\displaystyle -\frac{1}{2}V_{\alpha}(\bm{\xi}+\bm{\xi}_{*})\cdot\bm{\alpha},} & ({\displaystyle \varphi=\frac{1}{2}\bm{\xi}^{2})},
\end{cases}\label{eq:psi_p}
\end{equation}
\noindent
has been used \cite{CL88,F99}.
The above transformation gives rise to the concept of
the collisional part of the stress tensor $p_{ij}^\col$
and the heat-flow vector $q_i^\col$ for the transport in dense gases.
Therefore, in total, 
the stress tensor $p_{ij}$ and the heat-flow vector $q_i$ of the dense gas are expressed by
\begin{equation}
p_{ij}=p_{ij}^{\kin}+p_{ij}^{\col},\quad
q_i=q_i^{\kin}+q_i^{\col}. \label{eq:kin+col}
\end{equation}
\noindent
Note that, with these definitions,
the integration of \eqref{eq:MEE}
with respect to $\bxi$ after multiplied by $1$, $\xi_j$, and $(1/2)\bxi^2$ recovers the usual form of the conservation laws
\begin{subequations}
\begin{align}
& \frac{\partial \rho}{\partial t}
+\frac{\partial\rho v_i}{\partial X_i}=0, \\
& \frac{\partial \rho v_j}{\partial t}
+\frac{\partial}{\partial X_i}(\rho v_i v_j+p_{ij})=0, \\
& \frac{\partial}{\partial t}[\rho (e+\frac12\bm{v}^2)]
+\frac{\partial}{\partial X_i}
[\rho v_i (e+\frac12\bm{v}^2)+ p_{ij}v_j+q_i]=0.
\end{align}
\end{subequations}

\subsection{Relation to the equation of state\label{sec:Eos}}

Since the pressure $p$ is defined as the one third of the trace of the stress tensor, it is expressed as
\begin{multline}
p=\frac13 (p_{ii}^{\kin}+p_{ii}^{\col}) 
 =\rho RT
 +\frac{\sigma^{2}}{6m}\int {\int_{0}^{\sigma}}
   V_{\alpha}^{2}\theta(V_{\alpha}) g(\bm{X}_{\spa\bm{\alpha}}^{+},\bm{X}_{(\spa-\sigma)\bm{\alpha}}^{+})\\
   \times f_{*}(\bm{X}_{(\spa-\sigma)\bm{\alpha}}^{+})f(\bm{X}_{\spa\bm{\alpha}}^{+}){d\spa} d\Omega(\bm{\alpha})d\bm{\xi}_{*}d\bm{\xi}.
   \label{eq:pressure}
\end{multline}
This gives the connection to the EoS
that should be recovered at the uniform equilibrium state.

Consider now the infinite expanse of the dense gas in the uniform equilibrium state. 
In this case, $g$ is identical to $\sfg$,
and, as far as $\sfg$ takes the form of \eqref{eq:g_def} or \eqref{eq:gY},
the functions $f$, $f_*$, and $g$ are all independent of the position.
Accordingly, the last integral is reduced to
$(b\rho)\rho RT \sfg$ with $b=(2\pi/3)(\sigma^{3}/m)$,
which leads to the following EoS for the gas under consideration:
\begin{equation}
p=\rho RT (1+b\rho \sfg).
\end{equation}
In the present setting, since $\mR$ is computed from \eqref{eq:R_def} as
\begin{equation}
\mR
=\frac{\rho}{m}\int_{\Bbb{R}^3}  \theta(\sigma-|\bY-\bX|)d\bY
=\frac{\rho}{m}\frac{4\pi}{3}\sigma^3=2b\rho, \label{eq:mR_uniform}
\end{equation}
\noindent
the above $\sfg$ is expressed as
\begin{equation}
\sfg=
\begin{cases}
2{\mS}(\mR)=2{\mS}(2b\rho), \quad \mbox{(for EESM)},\\
 \mY(\rho),\quad \mbox{(for OEE)}.
\end{cases}\label{eq:sfg_uniform}
\end{equation}
\noindent
Hence, we may relate $\mS$ and $\mY$ for the same gas by
$\mY(\rho)=2\mS(2b\rho)$.

In closing the present subsection,
we present the specific form of ${\mS}$
for two typical EoS's in the literature:
the van der Waals equation of state \cite{vdW} for non-attractive molecules
\begin{equation}
p=\frac{\rho RT}{1-b\rho}=\rho RT(1+\frac{b\rho}{1-b\rho}),
\label{eq:EoSvdW}
\end{equation}
and the Carnahan--Starling equation of state \cite{CS69}
\begin{equation}
p=\rho RT\frac{1+\eta+\eta^2-\eta^3}{(1-\eta)^3}
 =\rho RT(1+\frac{4\eta-2\eta^2}{(1-\eta)^3}),
\label{eq:EoSCS}
\end{equation}
\noindent
where $\eta={b\rho}/{4}$.
It is readily seen that the appropriate form of ${\mS}$  
for the van der Waals equation of state for non-attractive molecules
is
\begin{equation}
{\mS}(x)=\frac{1}{2-x}, \label{eq:SvdW}
%\quad
%\int_0^x {\mS}(y) dy=-\ln (2-x)
\end{equation}
\noindent
while that for the Carnahan--Starling equation of state is
\begin{equation}
 {\mS}(x)=\frac{16(16-x)}{(8-x)^3}. \label{eq:SCS}
%\quad
%\int_0^x {\mS}(y) dy=-\ln (2-x)
\end{equation}

\begin{remark}
Note that
the collision term of the Enskog equation is not responsible 
for the attractive part of the EoS.
The attractive part is to be recovered by the Vlasov term of the Enskog--Vlasov equation \cite{G71,FGLS18}.
%Therefore the form \eqref{eq:SvdW} applies to the EESM
%with the Vlasov term for the \emph{full} version of the van der Waals fluids as well (see Appendix~\ref{app:EV}).
\end{remark}

\section{Fluid-dynamic limit\label{sec:FDlimit}}

\subsection{Dimensionless presentation\label{sec:dimensionless}}

In the present subsection, the original system of equations is made dimensionless
in order to figure out the embedded independent parameters.

Let $L$, $\rho_0$, and $T_0$ be the reference length, density, and temperature, respectively.
Let $\bm{x}=\bm{X}/L$, $\bz=\bm{\xi}/\sqrt{2RT_0}$,
and $\htt=t/(L/\sqrt{2RT_0})$,
which are, respectively, the dimensionless position vector, molecular velocity,
and time.
Then, the Enskog equation is recast as the counterpart
for the dimensionless VDF $\hf(\htt,\bx,\bz)=f(t,\bm{X},\bm{\xi})/[\rho_0(2RT_0)^{-3/2}]$,
i.e.,
\begin{subequations}\label{eq:hMEE}
\begin{align}
 & \frac{\partial \hf}{\partial \htt}+\z_{i}\frac{\partial \hf}{\partial x_{i}}
 =\frac{1}{k} \hJ(\hf)\equiv \frac{1}{k} [\hJ^{G}(\hf)-\hJ^{L}(\hf)],\quad \mathrm{for\ }\bx\in \hD,\displaybreak[0]\label{eq:h2.1}\\
 & \hJ^{G}(\hf)\equiv\frac{1}{2\sqrt{2\pi}}\int {\hg(\bx_{\hsg\bm{\alpha}}^{+},\bx)\hf_{*}^{\prime}(\bx_{\hsg\bm{\alpha}}^{+})\hf^{\prime}(\bx)}\hV_{\alpha}\theta(\hV_{\alpha})d\Omega(\bm{\alpha})d\bz_{*},\displaybreak[0]\label{eq:h2.2}\\
 & \hJ^{L}(\hf)\equiv\frac{1}{2\sqrt{2\pi}}\int {\hg(\bx_{\hsg\bm{\alpha}}^{-},\bx)\hf_{*}(\bx_{\hsg\bm{\alpha}}^{-})\hf(\bx)}\hV_{\alpha}\theta(\hV_{\alpha})d\Omega(\bm{\alpha})d\bz_{*},\label{eq:h2.3}
\end{align}
\end{subequations}
\noindent
where 
\noindent
\begin{subequations}
\begin{equation}\label{eq:Kn}
k=\frac{\sqrt{\pi}}{2}\frac{\ell_0}{L},\quad
\ell_0=\frac{1}{\sqrt{2}\pi\sigma^2(\rho_0/m)\sfg_0},
\end{equation}
\noindent
and $\sfg_0$ being $\sfg$ evaluated in the infinite expanse of the gas at a uniform density $\rho_0$, 
$\hsg=\sigma/L$,
$\hD$ is the dimensionless counterpart of $D$,
$\hg=g/\sfg_0$,  $\bx_{\bm{y}}^{\pm}=\bx\pm\bm{y}$,
and the following notation convention has been used:
\begin{align}
 & \begin{cases}
{
\hf(\bx)=\hf(\bx,\bz),\ \hf^{\prime}(\bx)=\hf(\bx,\bz^{\prime})},\\
{
\hf_{*}(\bx_{\hsg\bm{\alpha}}^{-})=\hf(\bx_{\hsg\bm{\alpha}}^{-},\bz_{*}),\ \hf_{*}^{\prime}(\bx_{\hsg\bm{\alpha}}^{+})=\hf(\bx_{\hsg\bm{\alpha}}^{+},\bz_{*}^{\prime})},
\end{cases}\label{eq:hcontf}\displaybreak[0]\\
 & \bz^{\prime}=\bz+\hV_{\alpha}\bm{\alpha},\quad\bz_{*}^{\prime}=\bz_{*}-\hV_{\alpha}\bm{\alpha},\quad \hV_{\alpha}=\hbV\cdot\bm{\alpha},\quad\hbV=\bz_{*}-\bz.\label{eq:h2.5}
\end{align}
\end{subequations}
\noindent
Note that $k$ is the Knudsen number $\Kn$ multiplied by $\sqrt{\pi}/2$ and $\ell_0$ is the reference mean free path of gas molecules which becomes identical to that of the dilute hard-sphere gas when $\sfg_0=1$; 
thus $\sfg_0$ represents the dense-gas effect on the reference mean free path.
Since $\sfg_0$ is $\sfg$ evaluated in the infinite expanse of the gas at a uniform density $\rho_0$, it is expressed by \eqref{eq:mR_uniform} and \eqref{eq:sfg_uniform} as
\begin{subequations}\label{eq:subsid}
\begin{equation}
 \sfg_0=2{\mS}({\mR}_0),\quad
 {\mR}_0=2b\rho_0,
\label{eq:g0mR0_def}
\end{equation}
\noindent
while $\hg$ and related dimensionless quantities are expressed as follows:
\begin{align}
& \hg(\bx,\bm{y})=[\hat{{\mS}}(\hmR(\bx))+\hat{{\mS}}(\hmR(\bm{y}))]\chi_\hD(\bx)\chi_\hD(\bm{y}),
\label{eq:hg_def}
\\
&\hat{{\mS}}(\hmR(\bx))
=\frac{{\mS}({\mR}(\bm{X}))}{\sfg_0}
=\frac12\frac{{\mS}({\mR}(\bm{X}))}{\mS(\mR_0)},
\label{eq:hS_def}
\end{align}%
\begin{align}
\hmR(\bx)
\equiv\frac{{\mR}(\bm{X})}{\mR_0}
=&\frac{\rho_0L^3}{m\mR_0}\int_{\hD} \hat{\rho}(\bm{y})\theta(\hsg-|\bm{y}-\bm{x}|) d\bm{y} \label{eq:hR_def}
\\
=&\big(\frac{4\pi\hsg^{3}}{3}\big)^{-1}
  \int_{\hD} \hat{\rho}(\bm{y})\theta(\hsg-|\bm{y}-\bm{x}|) d\bm{y},\notag
\end{align}%
\begin{equation}
 \hat{\rho}(\htt,\bx)[\equiv\rho(t,\bm{X})/\rho_0]=\int \hf(\htt,\bx,\bz) d\bz.
\label{eq:hr_def}
\end{equation}
\end{subequations}
\noindent
As is clear from \eqref{eq:hMEE}--\eqref{eq:subsid}, three dimensionless parameters $k$, $\hsg$, and ${\mR}_0$ are embedded in the Enskog equation.
Since $b/4=(\pi/6)(\sigma^3/m)$, 
${\mR}_0/8(=b\rho_0/4)$ is the volume fraction occupied by molecules 
of the gas with the density $\rho_0$.
Since we are concerned with the dense gas,
${\mR}_0$ is a finite positive constant.
The dilute gas limit is the case where ${\mR}_0$ tends to zero.
Since $\sfg_0=2{\mS}({\mR}_0)$ and the three parameters are related one another as
\begin{equation}
\hsg=\frac{3{\mR}_0\sfg_0}{\sqrt{2\pi}}k\equiv \kappa k,\label{eq:k-sigma}
\end{equation}
\noindent
$\hsg$ is proportional to $k$ when ${\mR}_0$ is fixed.
This is the key observation when discussing the fluid-dynamic limit.

In closing this subsection, we list the dimensionless version
of the macroscopic quantities for later convenience.
In addition to the dimensionless density $\hat{\rho}$ already given in \eqref{eq:hr_def},
the dimensionless flow velocity $\hbv=\bm{v}/\sqrt{2RT_0}$ (or $\hv_i$) and temperature $\hT=T/T_0$
are expressed as
\begin{subequations}
\begin{equation}
 \hv_i=\frac{1}{\hr}\langle \z_i \hf \rangle, \quad
 \hT=\frac{2}{3\hr}\langle \hbC^2 \hf \rangle, 
 \label{eq:vT_def}
\end{equation}
\noindent
and the dimensionless specific internal energy 
$\he=e/(RT_0)$,
the dimensionless kinetic part of the stress tensor $\hp_{ij}^{\kin}=p_{ij}^{\kin}/(\rho_0 RT_0)$, and that of the heat-flow vector $\hat{\bm{q}}^{\kin}={\bm{q}}^{\kin}/[(1/2)\rho_0(2RT_0)^{3/2}]$ (or $\hq_i^{\kin}$), are expressed as
\begin{equation}
 \he=\frac{1}{\hat{\rho}}\langle \hbC^2 \hf \rangle
  (=\frac32\hat{T}), \quad
 \hp_{ij}^{\kin}=2\langle \hC_i\hC_j\hf\rangle, \quad
 \hq_i^{\kin}=\langle\hC_i\hbC^2 \hf\rangle, 
 \label{eq:higher_mom_def}
\end{equation}
\end{subequations}
\noindent
where $\hbC=\bz-\bv$ (or $\hC_i=\z_i-\hv_i$) and $\langle\bullet\rangle=\int \bullet d\bz$.

\subsection{Scaling and situation\label{sec:scaling}}

Since we are concerned with the fluid-dynamic limit of the dense gas,
we fix $\kappa$ in \eqref{eq:k-sigma} as a constant of $O(1)$
and set $k\ll 1$ and $\hsg\ll 1$ as the same order small parameters.
Hereinafter, $k$ will be denoted by $\ve$, i.e.,
\begin{equation}
\ve\equiv k(=\frac{\sqrt{\pi}}{2}\frac{\ell_0}{L}),
\end{equation}
\noindent
to emphasize that it is a small parameter. 
Accordingly $\hsg$ is related to $\ve$ as $\hsg=\kappa\ve$.

\subsection{Grad's procedure of Chapman--Enskog method\label{sec:CE}}

We introduce $\psi_0=1$, $\psi_i=\z_i$ ($i=1,2,3$), and $\psi_4=\bz^2$
and their moments ${\vr}_s=\langle \psi_s \hf \rangle$ ($s=0,1,2,3,4$)
for the brevity of notation.
The $\vr$'s are expressed by the dimensionless density, velocity, and temperature
 (or specific internal energy) as $\vr_0=\hat{\rho}$, $\vr_i=\hat{\rho}\hv_i$ ($i=1,2,3$),
and $\vr_4=\hat{\rho}(\he+\hbv^2)=\hat{\rho}(\frac32\hT+\hbv^2)$.

In the Chapman--Enskog (CE) method,
the solution is sought in the form \cite{G58,S07}
\begin{equation}
\hf=\hf(\vr_r,\nbvr_r,\bz,\ve),
\label{eq:ansatz}
\end{equation}
\noindent
where $\nbvr_r$ represents partial spatial derivatives of $\vr_r$ of arbitrary orders
and $r=0,1,\dots,4$.
Thus, the $\psi_s$ moment of \eqref{eq:h2.1} ought to take the form
%
%\begin{subequations}
\begin{equation}
 \frac{\partial \vr_s}{\partial \htt}
+{\mF}_s(\vr_r,\nbvr_r,\ve)
=\frac{1}{\ve}{\mC}_s(\vr_r,\nbvr_r,\ve),\quad (s=0,\dots,4),
\label{eq:CE1}
\end{equation}
\noindent
with
\begin{equation}
\mF_s=\frac{\partial}{\partial x_i}\langle\psi_s\zeta_i \hf\rangle,
\quad
\mC_s=\langle\psi_s\hJ(\hf)\rangle,
\end{equation}
\noindent
which leads to the system of five partial differential equations 
for five unknown fluid-dynamic quantities $\vr_0,\cdots,\vr_4$.
This is the essential structure of the CE method.
The CE method further limits the solution
to the class that is expressed in the form
\begin{equation}
\hf=\sum_{n=0}^\infty \ve^n \hf^{(n)}(\vr_r,\nbvr_r,\bz).
\label{eq:f_exp}
\end{equation}
Since $\mF_s$ are expressed as
\begin{align}
  {\mF}_0
= &\frac{\partial}{\partial x_i}\langle \z_i\hf\rangle
= \frac{\partial\rho \hv_i}{\partial x_i}, 
\label{eq:CF0}\\
  {\mF}_j
= &\frac{\partial}{\partial x_i}\langle \z_i\z_j\hf\rangle
=  \frac{\partial}{\partial x_i}(\frac12\hp_{ij}^\kin+\hr \hv_i \hv_j), 
\label{eq:CFj}\\
  {\mF}_4
= &\frac{\partial}{\partial x_i}\langle \z_i\bz^2\hf\rangle
=  \frac{\partial}{\partial x_i}(\hr \hv_i (\he+\hbv^2)+\hp_{ij}^{\kin}\hv_j+\hq_i^{\kin}], 
\label{eq:CF4}
\end{align}
\noindent
the substitution of \eqref{eq:f_exp} yields the successive approximations
of $\hp_{ij}^{\kin}$ and $\hq_i^{\kin}$, i.e.,
\begin{equation}
\hp_{ij}^{\kin}=\hp_{ij}^{\kin(0)}+\hp_{ij}^{\kin(1)}\ve+\cdots, \quad
\hq_{i}^{\kin}=\hq_{i}^{\kin(0)}+\hq_{i}^{\kin(1)}\ve+\cdots,
\end{equation}
\noindent
with
\begin{equation}
\hp_{ij}^{\kin(n)}=2\langle \hC_i\hC_j\hf^{(n)}\rangle,
\quad
\hq_{i}^{\kin(n)}=\langle \hC_i\hbC^2\hf^{(n)}\rangle,
\end{equation}
\noindent
as in the case of the Boltzmann equation.
The main difference from the Boltzmann equation 
comes from the presence of $\mC_s$ and 
its contribution to the momentum and the energy transport.
Indeed, since $\mC_s$ can be expressed as well in the form:
\begin{subequations}\label{eq:mCs1}
\begin{align}
& \mC_0(\equiv\langle \hJ(\hf)\rangle)=0, \\
& \mC_j(\equiv\langle \z_j \hJ(\hf)\rangle)
       =-\frac12\ve\frac{\partial}{\partial x_i}\hp_{ij}^{\col}, \\
& \mC_4(\equiv\langle \bz^2\hJ(\hf)\rangle)
       =-\ve\frac{\partial}{\partial x_i}(\hp_{ij}^{\col}\hv_j+\hq_i^\col),
\end{align}
\end{subequations}
\noindent
with
\begin{subequations}\label{eq:pqcol}
\begin{align}
\hp_{ij}^{\col}= & \frac{1}{2\sqrt{2\pi}\ve}
\int {\int_{0}^{\hsg}}\alpha_{i}\alpha_{j}\hV_{\alpha}^{2}\theta(\hV_{\alpha})  \label{eq:col_stress_less}\\
&\qquad
 \times\hg(\bx_{\hspa\bm{\alpha}}^{+},\bx_{(\hspa-\hsg)\bm{\alpha}}^{+})\hf_{*}(\bx_{(\hspa-\hsg)\bm{\alpha}}^{+})\hf(\bx_{\hspa\bm{\alpha}}^{+}){d\hspa} d\Omega(\bm{\alpha})d\bz_{*}d\bz, \notag 
\\
\hq_{i}^{\col}= & \frac{1}{4\sqrt{2\pi}\ve}
\int{\int_{0}^{\hsg}}\alpha_{i}[(\hbC+\hbC_{*})\cdot\bm{\alpha}]\hV_{\alpha}^{2}\theta(\hV_{\alpha})
\label{eq:col_heat_less}
 \\
 & \qquad 
\times \hg(\bx_{\hspa\bm{\alpha}}^{+},\bx_{(\hspa-\hsg)\bm{\alpha}}^{+})
\hf_{*}(\bx_{(\hspa-\hsg)\bm{\alpha}}^{+})\hf(\bx_{\hspa\bm{\alpha}}^{+}){d\hspa} d\Omega(\bm{\alpha})d\bz_{*}d\bz,\notag 
\end{align}
\end{subequations}
\noindent
\eqref{eq:CE1} is reduced to the set of dimensionless conservation equations:
\begin{subequations}\label{eq:CSE}
\begin{align}
&
 \frac{\partial}{\partial\htt}\hr
+\frac{\partial}{\partial x_i}\hr\hv_i
=0, \\
&
 \frac{\partial}{\partial\htt}\hr\hv_j
+\frac{\partial}{\partial x_i}(\hr\hv_i\hv_j+\frac12\hp_{ij})
=0, \\
&
 \frac{\partial}{\partial\htt}\hr(\he+\hbv^2)
+\frac{\partial}{\partial x_i}[\hr\hv_i(\he+\hbv^2)+\hp_{ij}\hv_j+\hq_i]
=0,
\end{align}
with
\begin{align}
& \hp_{ij}=\hp_{ij}^\kin+\hp_{ij}^\col, \\
& \hq_{i}=\hq_{i}^\kin+\hq_{i}^\col.
\end{align}
\end{subequations}
\noindent
The approximations of $\hp_{ij}^\col$ and $\hq_i^{\col}$
as well as $\hJ(\hf)$ in terms of the expansion with respect to $\ve$
demand additional calculations
that are absent from the case of the Boltzmann equation.

In order to obtain such an approximation of $\hJ(\hf)$,
it is necessary first to expand it with respect to another small parameter $\hsg$, which results in the form
\begin{equation}
\hJ(\hf)=\mJ^{[0]}(\hf,\hf)+\hsg\mJ^{[1]}(\hf,\hf)+O(\hsg^2),
\label{eq:J_exp1}
\end{equation}
\noindent
where
\begin{subequations}\label{eq:Col_exp}
\begin{align}
 \mJ^{[0]}(F,G)
=&\mG
 \int (F_*^\prime G^\prime-F_*G) 
 \frac{\hV_\alpha\theta(\hV_\alpha)}{2\sqrt{2\pi}}d\Omega(\bm{\alpha})d\bz_*,  
 \label{eq:mJ0_def} \\
 \mJ^{[1]}(F,G)
=&\mG
 \int \alpha_i(\frac{\partial F_*^\prime}{\partial x_i}G^\prime+\frac{\partial F_*}{\partial x_i}G) \frac{\hV_\alpha\theta(\hV_\alpha)}{2\sqrt{2\pi}}d\Omega(\bm{\alpha})d\bz_* \label{eq:mJ1_def}\\
&+\frac12\frac{\partial \mG}{\partial x_i}
 \int \alpha_i(F_*^\prime G^\prime+F_*G) \frac{\hV_\alpha\theta(\hV_\alpha)}{2\sqrt{2\pi}}d\Omega(\bm{\alpha})d\bz_*
\notag\\
&+\frac{\hg^{[1]}}{\mG} \mJ^{[0]}(F,G), \notag
\end{align}
and
\begin{equation}
\mG(\bx)\equiv\lim_{\hsg\to0}\hg(\bx,\bx), \quad
\hg^{[1]}(\bx,\bx)\equiv\lim_{\hsg\to0}\partial \hg(\bx,\bx)/\partial \hsg.
\end{equation}
\end{subequations}
\noindent
Then, by substituting \eqref{eq:f_exp} and taking \eqref{eq:k-sigma} into account, 
\eqref{eq:J_exp1} is further transformed into the expansion 
with respect to $\ve$ as
\begin{subequations}
\begin{equation}\label{eq:hJ_exp_eps}
 \hJ(\hf)
=\mJ^{[0]}(\hf^{(0)},\hf^{(0)})
+[\kappa\mJ^{[1]}(\hf^{(0)},\hf^{(0)})+\mL(\hf^{(1)})]\ve+\cdots,
\end{equation}
with
\begin{equation}
 \mL(F)
=\mJ^{[0]}(F,\hf^{(0)})+\mJ^{[0]}(\hf^{(0)},F). \label{eq:mL_def}
\end{equation}
\end{subequations}
\noindent
In \eqref{eq:mJ0_def} and \eqref{eq:mJ1_def}, not only the time dependence but also the spatial dependence of functions have been suppressed from the notation, since it is no longer necessary to discriminate the difference of positions. 
The outline of derivation of \eqref{eq:Col_exp} is given in Appendix~\ref{app:ExpCol}.
The above expressions will be used to get the expressions of
$\hf^{(0)}$, $\hf^{(1)}$, and so on.

The same procedure applies as well to 
the expansions of $\hp_{ij}^{\col}$ and $\hq_i^{\col}$
and eventually yields 
\begin{align}
  \hp_{ij}^\col
=&\hp_{ij}^{\col(0)}+\ve\hp_{ij}^{\col(1)}+\cdots, \\
  \hq_{i}^\col
=&\hq_{i}^{\col(0)}+\ve\hq_{i}^{\col(1)}+\cdots, 
\end{align}
with
\begin{subequations}\label{eq:hpq_exps}
\begin{align}
\hp_{ij}^{\col(0)}
=&\frac{2}{15}\sqrt{\frac{\pi}{2}}\kappa\mG
\int (\hV_i\hV_j+\frac12\hbV^2\delta_{ij})
      \hf_*^{(0)}\hf^{(0)} d\bz_{*}d\bz, \\
\hq_{i}^{\col(0)}
= & \frac{1}{15}\sqrt{\frac{\pi}{2}}\kappa\mG
\int (\hC_j+\hC_{*j})
    (\hV_i\hV_j+\frac12\hbV^2\delta_{ij})
    \hf_*^{(0)}\hf^{(0)} d\bz_{*}d\bz,
\end{align}
and
\begin{align}
\hp_{ij}^{\col(1)}
=&\frac{2}{15}\sqrt{\frac{\pi}{2}}\kappa\mG
\int\Big\{
    (\hV_i\hV_j+\frac12\hbV^2\delta_{ij})
    (\hf_*^{(0)}\hf^{(1)}+\hf_*^{(1)}\hf^{(0)} 
 \label{eq:hp1}\\
  & +\frac{\kappa\hg^{[1]}}{\mG}\hf^{(0)}\hf_*^{(0)})+\frac{5}{32}\kappa
  \Big(
   \frac{\hV_i\hV_j\hV_k}{|\hbV|}+|\hbV|(\hV_i\delta_{jk}+\hV_j\delta_{ki}\notag\\
  &+\hV_k\delta_{ij})
  \Big) (\hf_*^{(0)}\frac{\partial \hf^{(0)}}{\partial x_k}
 -\hf^{(0)}\frac{\partial \hf_*^{(0)}}{\partial x_k})
  \Big\} d\bz_{*}d\bz, \notag\\
\hq_{i}^{\col(1)}
= & \frac{1}{15}\sqrt{\frac{\pi}{2}}\kappa\mG
\int (\hC_j+\hC_{*j})\Big\{
    (\hV_i\hV_j+\frac12\hbV^2\delta_{ij})
    (\hf_*^{(1)}\hf^{(0)}+\hf_*^{(0)}\hf^{(1)}
 \label{eq:hq1}\\
 & +\frac{\kappa\hg^{[1]}}{\mG}\hf^{(0)}\hf_*^{(0)}) +\frac{5}{32}\kappa
  \Big(
   \frac{\hV_i\hV_j\hV_k}{|\hbV|}+|\hbV|(\hV_i\delta_{jk}+\hV_j\delta_{ki}\notag\\
  &+\hV_k\delta_{ij})
  \Big) (\hf_*^{(0)}\frac{\partial \hf^{(0)}}{\partial x_k}
 -\hf^{(0)}\frac{\partial \hf_*^{(0)}}{\partial x_k})
  \Big\} d\bz_{*}d\bz.\notag
\end{align}
\end{subequations}
\noindent
The derivation of \eqref{eq:hpq_exps} is given in Appendix~\ref{app:hpq_col}.

\begin{remark}
Since the specific form of $\mG$ is common between the EESM and the OEE
for the common equation of state
(see Appendix~\ref{app:g_specf}),
the EESM recovers the same fluid-dynamic system 
as that derived from the OEE at least up to $O(\ve)$ inclusive.
\end{remark}
\begin{remark}
As will be seen soon later,
in evaluating $O(\ve)$ term of \eqref{eq:hJ_exp_eps},
the last term on the right-hand side of \eqref{eq:mJ1_def}
vanishes regardless of $\hg^{[1]}$ thanks to \eqref{eq:mJ0_sol} below.
The term of $\hg^{[1]}$ in \eqref{eq:hq1} vanishes as well regardless of $\hg^{[1]}$.
The information of the specific form of $\hg^{[1]}$ is required later
when reducing the expression \eqref{eq:hp1} to the final form \eqref{eq:hp_col1_final}. 
\end{remark}

With the above preparations,
let us turn to the procedure of approximation.
Since the substitution of \eqref{eq:f_exp} into \eqref{eq:hMEE}
gives rise to a series of equations for $\hf^{(n)}$:
\begin{align}
& \mJ^{[0]}(\hf^{(0)},\hf^{(0)})=0, \label{eq:mJ0_sol}\\
& \mL(\hf^{(1)})+\kappa\mJ^{[1]}(\hf^{(0)},\hf^{(0)})
=\frac{\partial\hf^{(0)}}{\partial \htt}+\zeta_i\frac{\partial\hf^{(0)}}{\partial x_i}, \label{eq:mL_sol}
\end{align}
and so on, $\hf^{(n)}$ can be determined successively from the lowest order.
Here it should be noticed that $\mJ^{[0]}$ and $\mL$ are respectively the collision operators of the original and the linearized Boltzmann equation multiplied by $\mG$ for a hard-sphere gas. 
Thus, we may simply follow the solution procedure for the Boltzmann equation.

Indeed, by \eqref{eq:mJ0_sol}, 
$\hf^{(0)}$ is identified as the local Maxwellian
\begin{equation}
\hf^{(0)}=\frac{\hr}{(\pi\hT)^{3/2}}\exp(-\frac{(\bz-\hbv)^2}{\hT}).
\label{eq:hf0}
\end{equation}
\noindent
Here are two observations.
Firstly, this form is consistent with the ansatz \eqref{eq:ansatz},
since $\hr$, $\hbv$, and $\hT$ are all expressed 
in terms of $\vr$'s.
Secondly, it is readily seen that
\begin{align}
\hp_{ij}^{\kin(0)}=&\frac23\langle \hbC^2 \hf^{(0)}\rangle\delta_{ij}=\hr\hT\delta_{ij}, \\
\hp_{ij}^{\col(0)}
=&\frac{1}{3}\sqrt{\frac{\pi}{2}}\kappa\hr\mG \hr\hT\delta_{ij}
\big(=\mR_0\mS(\mR_0)\hr\mG \hr\hT\delta_{ij}\big), \\
\hq_{i}^{\kin(0)}
= &\hq_{i}^{\col(0)}
= 0,
\end{align}
\noindent
so that
\begin{subequations}\label{eq:hpq_Euler}
\begin{align}
\hp_{ij}=&\hr\hT(1+\frac{1}{3}\sqrt{\frac{\pi}{2}}\kappa\hr\mG )\delta_{ij}+O(\ve), \\
\hq_{i}=&O(\ve).
\end{align}
\end{subequations}
Next, in order to improve the approximation, 
we try to find $\hf^{(1)}$ by solving \eqref{eq:mL_sol}.
To this end,
we start with the consideration of its consistency condition, namely
\begin{equation}
 \langle \psi_s \mL(\hf^{(1)})\rangle
=\langle\psi_s [\frac{\partial \hf^{(0)}}{\partial \htt}+\z_{i}\frac{\partial \hf^{(0)}}{\partial x_{i}}-\kappa \mJ^{[1]}(\hf^{(0)},\hf^{(0)})]\rangle,
\end{equation}
\noindent
that is reduced to
\begin{equation}
 \kappa\langle \psi_s\mJ^{[1]}(\hf^{(0)},\hf^{(0)}) \rangle
=\langle\psi_s (\frac{\partial \hf^{(0)}}{\partial \htt}+\z_{i}\frac{\partial \hf^{(0)}}{\partial x_{i}})\rangle,
\label{eq:solv_cond}
\end{equation}
\noindent
because of $\langle\psi_s\mL(F)\rangle=0$ for any $F$.
If \eqref{eq:solv_cond} does not hold, 
the solution of \eqref{eq:mL_sol} is not possible. 
The key point in the CE method is that 
\eqref{eq:solv_cond} gives rise to the relation 
that converts the time derivative of $\vr$'s
into their spatial derivatives
\begin{equation}
 \frac{\partial \vr_s}{\partial \htt}
=\mC_s^{(1)}(\vr_r,\nabla\vr_r)-\mF_s^{(0)}(\vr_r,\nabla\vr_r),
\label{eq:redConst}
\end{equation}
\noindent
so that the ansatz \eqref{eq:ansatz} is kept ensured,
where $\mF_s^{(0)}$ and $\mC_s^{(1)}$ are respectively
the zero-th and the 1st order expansion coefficient functions
of $\mF_s$ and $\mC_s$ with respect to $\ve$ and are given as
\begin{subequations}\label{eq:mFs}
\begin{align}
  \mF_0^{(0)}
 =&\frac{\partial}{\partial x_i}(\hr\hv_i),\\
 \mF_j^{(0)}
 =&\frac{\partial}{\partial x_i}(\hr\hv_i\hv_j+\frac12\hp_{ij}^{\kin(0)})
 =\frac{\partial}{\partial x_i}(\hr\hv_i\hv_j+\frac12\hr\hT\delta_{ij}), \\
 \mF_4^{(0)}
 =& \frac{\partial}{\partial x_i}[\hr\hv_i(\he+\hbv^2)+\hp_{ij}^{\kin(0)}\hv_j+\hq_i^{\kin(0)}] \\
 =& \frac{\partial}{\partial x_i}[\hr\hv_i(\he+\hbv^2)+\hr\hT\hv_i],\notag
\end{align}
\end{subequations}
\noindent
and
\begin{subequations}\label{eq:mCs}
\begin{align}
\mC_0^{(1)}
=&0,\\
\mC_j^{(1)}
=&-\frac12\frac{\partial}{\partial x_i}\hp_{ij}^{\col(0)}
 =-\frac{\kappa}{6}\sqrt{\frac{\pi}{2}} \frac{\partial(\hr^2\mG\hT)}{\partial x_j},\\
\mC_4^{(1)}
=&-\frac{\partial}{\partial x_i}(\hp_{ij}^{\col(0)}\hv_j+\hq_i^{\col(0)})
 =-\frac{\kappa}{3}\sqrt{\frac{\pi}{2}}\frac{\partial}{\partial x_i}
   (\hr^2\mG\hT\hv_i).
\end{align}
\end{subequations}

Using the relation \eqref{eq:redConst},
the right-hand side of \eqref{eq:mL_sol} 
can be transformed as
\begin{multline}
 \frac{\partial\hf^{(0)}}{\partial\htt}
+\z_i\frac{\partial\hf^{(0)}}{\partial x_i}
=\frac{\partial\hf^{(0)}}{\partial\vr_s}
 (\frac{\partial\vr_s}{\partial\htt}+\z_i\frac{\partial\vr_s}{\partial x_i})
=(\mC_s^{(1)}-\mF_s^{(0)}+\z_i\frac{\partial\vr_s}{\partial x_i})\frac{\partial\hf^{(0)}}{\partial\vr_s} \notag\\
=\Big[
\frac{2}{\hT} (\hC_i\hC_j-\frac13\hbC^2\delta_{ij}){\frac{\partial \hv_i}{\partial x_j}}+\frac{\hC_i}{\hT}(\frac{\hbC^2}{\hT}-\frac52)\frac{\partial\hT}{\partial x_i}
  \Big]\hf^{(0)}+\mC_s^{(1)}\frac{\partial\hf^{(0)}}{\partial\vr_s},
\end{multline}
\noindent
and thus \eqref{eq:mL_sol} is rewritten in the form
\begin{align}
 \mL(\hf^{(1)})
=&\Big[
\frac{2}{\hT} (\hC_i\hC_j-\frac13\hbC^2\delta_{ij}){\frac{\partial \hv_i}{\partial x_j}}+\frac{\hC_i}{\hT}(\frac{\hbC^2}{\hT}-\frac52)\frac{\partial\hT}{\partial x_i}
  \Big]\hf^{(0)} \label{eq:mLhf1}\\
& +\mC_s^{(1)}\frac{\partial\hf^{(0)}}{\partial\vr_s}
  -\kappa\mJ^{[1]}(\hf^{(0)},\hf^{(0)}).
  \notag
\end{align}

The last two terms on the right-hand side
are peculiar to the Enskog equation
and have to be newly evaluated.
Details of the calculations are summarized in Appendices~\ref{app:Inhom}
and \ref{app:mJ1_hf0_hf0}.
The results eventually obtained are as follows:
\begin{align}
   \mC_s^{(1)}\frac{\partial\hf^{(0)}}{\partial\vr_s}
=&-\frac{\kappa}{3}\sqrt{\frac{\pi}{2}}
   \frac{\partial(\hr\mG)}{\partial x_j}
  \hC_j\hf^{(0)}
  -\frac{\kappa}{3}\sqrt{\frac{\pi}{2}}
  \hr\mG \frac{\partial\ln(\hr\hT)}{\partial x_j}
  \hC_j\hf^{(0)}\\
&-\frac{\kappa}{3}\sqrt{\frac{\pi}{2}}
  \hr\mG
  \frac23\frac{\partial\hv_j}{\partial x_j}(\frac{\hbC^2}{\hT}-\frac32)\hf^{(0)},
\notag\end{align}
\noindent
and
\begin{align}
% &
 \mJ^{[1]}(\hf^{(0)},\hf^{(0)})
 % \notag \\
=&-\frac13\sqrt{\frac{\pi}{2}} \hr\mG\hf^{(0)}
  \Big[\frac{4}{5}(\frac{\hC_i\hC_j}{\hT}
                  +\frac12(\frac{\hbC^2}{\hT}-\frac52)\delta_{ij})
                  \frac{\partial\hv_j}{\partial x_i}
 \\
&+\frac{3}{5}\hC_i(\frac{\hbC^2}{\hT}-\frac52)
  \frac{\partial}{\partial x_i}\ln\hT 
 +\hC_i\frac{\partial\ln(\hr^2\mG\hT)}{\partial x_i}\Big].\notag
\end{align}
\noindent
Combining these together eventually yields
\begin{align}
& \mC_s^{(1)}\frac{\partial\hf^{(0)}}{\partial\vr_s}
 -\kappa\hJ^{[1]}(\hf^{(0)},\hf^{(0)})  \\
=&\frac{\kappa}{5}\sqrt{\frac{\pi}{2}}\hr\mG \hf^{(0)}
  \Big[\frac{4}{3}
  \frac{1}{\hT}(\hC_i\hC_j-\frac13\hbC^2\delta_{ij})
   \frac{\partial\hv_j}{\partial x_i}
   +\frac{\hC_i}{\hT}(\frac{\hbC^2}{\hT}-\frac52)
  \frac{\partial\hT}{\partial x_i}\Big],\notag
\end{align}
\noindent
so that \eqref{eq:mL_sol} is finally reduced to
the following equation for $\hf^{(1)}$:
\begin{align}
 \mL(\hf^{(1)})
=&\Big[
\frac{2}{\hT} (\hC_i\hC_j-\frac13\hbC^2\delta_{ij})
( 1+\frac{2\kappa}{15}\sqrt{\frac{\pi}{2}} \hr\mG )
{\frac{\partial \hv_i}{\partial x_j}}\label{eq:mL_final}
   \\
& +\frac{\hC_i}{\hT}(\frac{\hbC^2}{\hT}-\frac52)
   (1+\frac{\kappa}{5}\sqrt{\frac{\pi}{2}}\hr\mG )
   \frac{\partial\hT}{\partial x_i}
\Big]\hf^{(0)}.\notag
\end{align}
\noindent
Remind that 
$\mL$ is identical to the linearized Boltzmann collision operator
multiplied by $\mG$.
Thus, \eqref{eq:mL_final} can be solved in the same way as in the case of the Boltzmann equation to yield
\begin{align}
 \hf^{(1)}
=&\frac{1}{\hr\mG\sqrt{\hT}}\Big[- B(\frac{|\hbC|}{\sqrt{\hT}})
\frac{1}{\hT} (\hC_i\hC_j-\frac13\hbC^2\delta_{ij})
( 1+\frac{2}{5}\Lambda )
{\frac{\partial \hv_i}{\partial x_j}}\label{eq:hf1_final}
   \\
& -\frac{\hC_i}{\hT}A(\frac{|\hbC|}{\sqrt{\hT}})
   (1+\frac{3}{5}\Lambda)
   \frac{\partial\hT}{\partial x_i}
\Big]\hf^{(0)},\notag
\end{align}
\noindent
where $\Lambda=(1/3)\sqrt{\pi/2}\kappa\hr\mG$, while $A$ and $B$ are the solutions of
\begin{align}
& \mL[(\hC_i\hC_j-\frac13\hbC^2\delta_{ij})B(\frac{|\hbC|}{\sqrt{\hT}})\hf^{(0)}]
=-2(\hC_i\hC_j-\frac13\hbC^2\delta_{ij})\hr\mG\sqrt{\hT}\hf^{(0)}, \\
& \mL[\hC_iA(\frac{|\hbC|}{\sqrt{\hT}})\hf^{(0)}]
=-\hC_i(\frac{\hbC^2}{\hT}-\frac52)\hr\mG\sqrt{\hT}\hf^{(0)},
\end{align}
\noindent
with the subsidiary condition:
%\begin{equation}
$\int_0^\infty Z^4A(Z)\exp(-Z^2)dZ=0$.
%\end{equation}

Since $\hf^{(1)}$ has been obtained, we are ready to evaluate
$p_{ij}^{\kin(1)}$, $q_{i}^{\kin(1)}$, $p_{ij}^{\col(1)}$, and $q_{i}^{\col(1)}$.
The details of calculations are given in Appendices~\ref{app:pq_kin1} and \ref{app:pq_col1},
and only the results are presented here, i.e.,
\begin{subequations}\label{eq:hpq1-kin-col}
\begin{align}
 \hp_{ij}^{\kin(1)}
=&2\int \hC_i\hC_j\hf^{(1)}d\bz 
=-\frac{1}{\mG}( 1+\frac{2}{5}\Lambda )
   \gamma_1\sqrt{\hT}\overline{\frac{\partial \hv_i}{\partial x_j}},\\
 \hq_i^{\kin(1)}
=&\int\hC_i\hbC^2\hf^{(1)}d\bz
= -\frac{1}{\mG}(1+\frac{3}{5}\Lambda )
    \frac{5}{4}\gamma_2\sqrt{\hT}
   \frac{\partial\hT}{\partial x_i},
\end{align}
and
\begin{align}        
    \hp_{ij}^{\col(1)}
%  = & \frac{2}{15}\sqrt{\frac{\pi}{2}}\kappa\hr\mG
%    \hp_{ij}^{\kin(1)} 
%    -\frac{2}{15}\sqrt{\hT}\kappa^2\hr^2\mG
%     (\overline{\frac{\partial\hv_i}{\partial x_j}}
%     +\frac53\frac{\partial\hv_k}{\partial x_k}\delta_{ij})
%    \\
  = & \frac{2}{5}\Lambda \hp_{ij}^{\kin(1)} 
    -\frac{12}{5\pi}\frac{\sqrt{\hT}}{\mG}\Lambda^2
     \overline{\frac{\partial\hv_i}{\partial x_j}}
    -\frac{4}{\pi}\frac{\sqrt{\hT}}{\mG}\Lambda^2
     \frac{\partial\hv_k}{\partial x_k}\delta_{ij},\label{eq:hp_col1_final}
    \\
   \hq_{i}^{\col(1)}
=& \frac{3}{5}\Lambda \hq_i^{\kin(1)}
  -\frac{3}{\pi}\frac{\Lambda^2}{\mG}\sqrt{\hT}
   \frac{\partial \hT}{\partial x_i},
   \label{eq:hq_col1final}
\end{align}
\noindent
where $\overline{A_{ij}}=A_{ij}+A_{ji}-(2/3)A_{kk}\delta_{ij}$
and $\gamma_1$ and $\gamma_2$ are the following constants
\begin{align}
& \gamma_1\equiv\frac{8}{15\sqrt{\pi}}\int_0^\infty Z^6 B(Z)\exp(-Z^2)dZ\fallingdotseq 1.270042427,\\
& \gamma_2\equiv\frac{16}{15\sqrt{\pi}}\int_0^\infty Z^6 A(Z)\exp(-Z^2)dZ\fallingdotseq 1.922284066.
\end{align}
\end{subequations}
Therefore, we have obtained better approximations than \eqref{eq:hpq_Euler} as
\begin{subequations}\label{eq:hpq_final}
\begin{align}
   \hp_{ij}
=& \hr\hT(1+\Lambda)\delta_{ij}
  -\frac{\ve}{\mG}
   \Big[( 1+\frac{2}{5}\Lambda )^2
           +\frac{12}{5\pi\gamma_1}\Lambda^2
   \Big]\gamma_1\sqrt{\hT}\overline{\frac{\partial\hv_i}{\partial x_j}} \\
 &
       -\frac{4}{\pi}\frac{\ve}{\mG}\Lambda^2\sqrt{\hT}
        \frac{\partial\hv_k}{\partial x_k}\delta_{ij}+O(\ve^2), \notag\displaybreak[0]\\
   \hq_{i}
=&-\frac{\ve}{\mG}\Big[(1+\frac{3}{5}\Lambda )^2
  +\frac{12}{5\pi}\frac{\Lambda^2}{\gamma_2}
   \Big]\frac{5}{4}\gamma_2\sqrt{\hT}
   \frac{\partial \hT}{\partial x_i} +O(\ve^2).
\end{align}
\end{subequations}

Going back to \eqref{eq:CE1} or \eqref{eq:CSE}
and substituting \eqref{eq:hpq_final},
we recover the Euler set of equations
\begin{subequations}\label{eq:Euler}
\begin{align}
& \frac{\partial \hr}{\partial \htt}+\frac{\partial \hr \hv_i}{\partial x_{i}}=0, \displaybreak[0]\\
&\frac{\partial \hr\hv_j}{\partial \htt}
+\frac{\partial }{\partial x_{i}}(\hr\hv_j\hv_i +\frac12\hp\delta_{ij})
=0, \displaybreak[0]\\
& \frac{\partial \hr(\he+\hbv^2)}{\partial \htt}
+ \frac{\partial}{\partial x_i}[\hr (\he+\hbv^2) \hv_i +\hp\hv_i]
=0,
\end{align}
\end{subequations}
\noindent
coupled with $\he=(3/2)\hT$ and the EoS for a non-ideal gas
\begin{equation}\label{eq:EoS}
\hp=\hr\hT(1+\Lambda),\quad 
\Lambda=\frac13\sqrt{\frac{\pi}{2}}\kappa\hr\mG
\Big(
=\frac{\sqrt{2\pi}}{3}\kappa\hr\hmS(\hr)
\Big),
\end{equation}
with the error of $O(\ve)$,
while we recover the Navier--Stokes--Fourier (NSF) set of equations 
\begin{subequations}\label{eq:NS}
\begin{equation}
\frac{\partial \hr}{\partial \htt}+\frac{\partial \hr \hv_i}{\partial x_{i}}=0, 
\end{equation}%
\begin{equation}
 \frac{\partial \hr\hv_j}{\partial \htt}
+\frac{\partial }{\partial x_{i}}(\hr\hv_j \hv_i +\frac12\hp\delta_{ij})
=\frac12\frac{\partial}{\partial x_i}
  \Big(
         \hmu\overline{\frac{\partial\hv_i}{\partial x_j}}
        +\hmu_B\frac{\partial\hv_k}{\partial x_k}\delta_{ij}
  \Big),
\end{equation}%
\begin{multline}
  \frac{\partial \hr(\he+\hbv^2)}{\partial \htt}
+ \frac{\partial}{\partial x_i}[\hr(\he+\hbv^2) \hv_i +\hp\hv_i] \\
  =\frac{\partial}{\partial x_i}
   \Big(
       [\hmu\overline{\frac{\partial\hv_i}{\partial x_j}}
       +\hmu_B\frac{\partial\hv_k}{\partial x_k}\delta_{ij}]\hv_j
       +\hlam\frac{\partial \hT}{\partial x_i}
   \Big),
\end{multline}
\end{subequations}
\noindent
again coupled with $\he=(3/2)\hT$ and \eqref{eq:EoS}, where
\begin{subequations}
\begin{align}
   \hmu
=& \frac{\ve}{\mG}
   \Big[( 1+\frac{2}{5}\Lambda )^2
           +\frac{12}{5\pi\gamma_1}\Lambda^2
   \Big]\gamma_1\sqrt{\hT}, \displaybreak[0]\\
   \hmu_B
=& \frac{4}{\pi}\frac{\ve}{\mG}\Lambda^2\sqrt{\hT},\displaybreak[0] \\
   \hlam
=& \frac{\ve}{\mG}\Big[(1+\frac{3}{5}\Lambda )^2
  +\frac{12}{5\pi}\frac{\Lambda^2}{\gamma_2}
   \Big]\frac{5}{4}\gamma_2\sqrt{\hT},
\end{align}
\end{subequations}
\noindent
with the error of $O(\ve^2)$.
%The above $\hmu$, $\hmu_B$, and $\hlam$ are respectively the dimensionless
%viscosity, bulk viscosity, and thermal conductivity.

Finally, we close this subsection by recasting the system \eqref{eq:NS} in the dimensional form:
\begin{subequations}\label{eq:NS_dim}
\begin{equation}
\frac{\partial \rho}{\partial t}+\frac{\partial \rho v_i}{\partial X_{i}}=0, \label{eq:mass_dim}
\end{equation}%
\begin{equation}
 \frac{\partial \rho v_j}{\partial t}
+\frac{\partial }{\partial X_{i}}(\rho v_j v_i + p\delta_{ij})
=\frac{\partial}{\partial X_i}
  \Big(
         \mu\overline{\frac{\partial v_i}{\partial X_j}}
        +\mu_B\frac{\partial v_k}{\partial X_k}\delta_{ij}
  \Big),\label{eq:mom_dim}
\end{equation}%
\begin{multline}
  \frac{\partial}{\partial t} [\rho(e+\frac12\bv^2)]
+ \frac{\partial}{\partial X_i}[\rho(e+\frac12\bv^2) v_i + p v_i] \\
  =\frac{\partial}{\partial X_i}
   \Big(
       [\mu\overline{\frac{\partial v_i}{\partial X_j}}
       +\mu_B\frac{\partial v_k}{\partial X_k}\delta_{ij}] v_j
       +\lambda\frac{\partial T}{\partial X_i}
   \Big),\label{eq:en_dim}
\end{multline}
\end{subequations}
\noindent
coupled with $e=(3/2)RT$ and $p=\rho RT(1+\Lambda)$, where
\begin{subequations}\label{eq:tranpCoeff}
\begin{align}
&\Lambda\Big(=\frac13\sqrt{\frac{\pi}{2}}\kappa\hr\mG
        \Big)
             =2b\rho\mS(2b\rho),\displaybreak[0]\\
&  \mu
  \Big(=L\rho_0\sqrt{\frac{RT_0}{2}}\hmu\Big)
= \frac{1}{2\mS(2b\rho)}
   \Big[( 1+\frac{2}{5}\Lambda )^2
           +\frac{12}{5\pi\gamma_1}\Lambda^2
   \Big]\mu_\mathrm{dilute}(T), \displaybreak[0]\\
&  \mu_B
  \Big(=L\rho_0\sqrt{\frac{RT_0}{2}}\hmu_B\Big)
= \frac{1}{2\mS(2b\rho)}
  \frac{4}{\pi\gamma_1}\Lambda^2 \mu_\mathrm{dilute}(T),\displaybreak[0]\\
&  \lambda
  \Big(=\frac{\rho_0}{2}(2RT_0)^{3/2}\frac{L}{T_0}\hat{\lambda}\Big)
=  \frac{1}{2\mS(2b\rho)}\Big[(1+\frac{3}{5}\Lambda )^2
  +\frac{12}{5\pi}\frac{\Lambda^2}{\gamma_2}
   \Big]\lambda_\mathrm{dilute}(T),
\end{align}
\end{subequations}
\noindent
with $\mu_\mathrm{dilute}(T)$ and $\lambda_\mathrm{dilute}(T)$ being the viscosity and the thermal conductivity 
in the dilute gas limit,
i.e.,
\begin{equation}
\mu_\mathrm{dilute}(T)=\frac{\gamma_1}{4\sqrt{\pi}}\frac{m}{\sigma^2}\sqrt{RT},
\quad
\lambda_\mathrm{dilute}(T)=\frac{5R}{2}\frac{\gamma_2}{\gamma_1}\mu_\mathrm{dilute}(T).
\end{equation}
\noindent
The $\mu$, $\mu_B$, and $\lambda$ in \eqref{eq:tranpCoeff} are respectively the viscosity, bulk viscosity, and thermal conductivity of the gas.
The dimensional version of the Euler system \eqref{eq:Euler} is obtained 
by dropping the right-hand-side terms of \eqref{eq:mom_dim} and \eqref{eq:en_dim} from the system \eqref{eq:NS_dim}.
\section{Concluding remarks\label{sec:Conclusion}}

In the present paper, the fluid-dynamic limit of the EESM has been discussed
on the basis of Grad's procedure of the Chapman--Enskog expansion.
The EESM differs from the OEE in the correlation factor and ensures the H-theorem.
The key observation in the discussion of the fluid-dynamic limit for the dense gas
is the need to keep
the molecular diameter and the mean free path of the same order
when letting them tend to zero.
Although the procedure for the fluid-dynamic limit 
requires the same level of cumbersome manipulation as in the case of the OEE, 
it is along the same lines as the procedure in the case of the Boltzmann equation.
As expected, it has been shown that, depending on the accuracy of the approximation of the stress tensor and the heat-flow vector, the Euler and NSF sets of equations are obtained with the error of the Knudsen number and the square of the Knudsen number, respectively. It has also been shown that these fluid-dynamic sets are identical to those obtained by the OEE, including the specific form of the transport coefficients occurring in the NSF set.
Since the OEE is known to recover the fluid-dynamical transport properties well, the present results mean that the EESM provides consistent descriptions both thermodynamically and fluid-dynamically.

In the Enskog equation, only the repulsive interaction of molecules in the non-ideal gas effect is considered. In order to include the attractive interaction of molecules, an extension to the case of the Enskog--Vlasov equation is required. In this case, the internal energy as well as the stress tensor and the heat flow vector have a contribution from the Vlasov term, which represents the attractive interaction of the molecules. We will leave this extension for another occasion.

\section*{Acknowledgement}
The first author is a JSPS Research Fellowship DC1 and his research is supported by the JSPS Grant-in-Aid for JSPS fellows (No. 24KJ1450).
The authors thank Masanari Hattori for reading the manuscript and for helpful comments.

\appendix
\section{Expansion of the collision term\label{app:ExpCol}}

The collision term contains the small parameter $\hsg$.
Therefore, we will first consider the expansions of the integrand of the collision term with respect to $\hsg$, although $\ve$ and $\hsg$ are related to each other by $\ve=\kappa\hsg$. 
The expansion of $\hf$ with respect to $\ve$ will be treated later.

Before starting the procedure, it should be noted that
the $\vr_s$ and $\nabla\vr_s$ are treated as the same order quantities
in the CE method.
Hence, the method is applicable to the spatial region
away from the boundary.
This implies that we do not have to worry about possible truncations of the integral domain near the boundary. This situation makes the required procedure of expansions straightforward.

Keeping the above observation in mind,
first carry out the following expansion with respect to $\hsg$
\begin{subequations}\label{eq:sigma_expansion}
\begin{align}
\hg(\bx,\bx_{\hsg\bm{\alpha}}^\pm)
=&\hg(\bx,\bx)\pm\hsg\alpha_i\frac{\partial \hg(\bx,\bm{y})}{\partial y_i}\big|_{\bm{y}=\bx}+O(\hsg^2) \\
=&\hg^{[0]}(\bx,\bx)+\hsg\Big(\hg^{[1]}(\bx,\bx)\pm\alpha_i\frac{\partial \hg^{[0]}(\bx,\bm{y})}{\partial y_i}\big|_{\bm{y}=\bx}\Big)+O(\hsg^2),\notag\\
\hf_*^\prime(\bx_{\hsg\bm{\alpha}}^+)
=&\hf_*^\prime(\bx)+\hsg\alpha_i\frac{\partial \hf_*^\prime(\bm{x})}{\partial x_i}+O(\hsg^2),\\
\hf_*(\bx_{\hsg\bm{\alpha}}^-)
=&\hf_*(\bx)-\hsg\alpha_i\frac{\partial \hf_*(\bm{x})}{\partial x_i}+O(\hsg^2),
\end{align}
\end{subequations}
where the parameter dependence of the form of $\hg$ itself has been taken into account by the expansion $\hg(\bx,\by)=\hg^{[0]}(\bx,\by)+\hsg\hg^{[1]}(\bx,\by)+\cdots$.
Note that $\hg^{[0]}(\bx,\by)=\lim_{\hsg\to0}\hg(\bx,\by)$ 
and $\hg^{[1]}(\bx,\by)=\lim_{\hsg\to0}\partial\hg(\bx,\by)/\partial\hsg$.
If a new notation $\mG(\bx)\equiv\hg^{[0]}(\bx,\bx)$ is introduced,
\begin{equation}
  \frac{\partial \mG(\bx)}{\partial x_i}
=2\frac{\partial \hg^{[0]}(\bx,\by)}{\partial y_i}\big|_{\by=\bx},
\end{equation}
because of the symmetry $\hg^{[0]}(\bx,\by)=\hg^{[0]}(\by,\bx)$.
Then, using these notation and relation, it is readily seen that
\begin{multline}
\hg(\bx_{\hsg\bm{\alpha}}^+,\bx)
\hf_*^\prime(\bx_{\hsg\bm{\alpha}}^+) \\
=[\mG(\bx)+\hsg\Big(\hg^{[1]}(\bx,\bx)+\frac12\alpha_i\frac{\partial \mG(\bx)}{\partial x_i}\Big)+O(\hsg^2)]\\
\times [\hf_*^\prime(\bx)+\hsg\alpha_i\frac{\partial \hf_*^\prime(\bm{x})}{\partial x_i}+O(\hsg^2)] \\
=\mG(\bx)\hf_*^\prime(\bx)
 +\hsg\Big[\Big(\hg^{[1]}(\bx,\bx)+\frac12\alpha_i\frac{\partial \mG(\bx)}{\partial x_i}\Big)\hf_*^\prime(\bx) \\
+\mG(\bx)\alpha_i\frac{\partial \hf_*^\prime(\bm{x})}{\partial x_i}
\Big]+O(\hsg^2).
\end{multline}
A similar expression can be obtained for $\hg(\bx_{\hsg\bm{\alpha}}^-,\bx)
\hf_*(\bx_{\hsg\bm{\alpha}}^-)$.
Then, the substitution into $\hJ(\hf)$ leads to the expansion \eqref{eq:J_exp1} 
with the form of $\mJ^{[0]}$ and $\mJ^{[1]}$ given by \eqref{eq:mJ0_def} and \eqref{eq:mJ1_def}.

%Finally, substitution of \eqref{eq:f_exp} and the relation $\hsg=\kappa\ve$
%yield the expansion of $\hJ(\hf)$
%in the form \eqref{eq:hJ_exp_eps} with \eqref{eq:Col_exp}.

\section{Expansions of $\hp_{ij}^\col$ and $\hq_i^\col$ \label{app:hpq_col}}

We will derive the expansions of $\hp_{ij}^{\col}$ and $\hq_{i}^{\col}$ 
on the basis of the expression \eqref{eq:pqcol}.
Since $\hspa$ varies between $0$ and $\hsg$ in \eqref{eq:pqcol}, 
we can use the following expansion, which is similar to that in Appendix~\ref{app:ExpCol}:
\begin{align}
&\hg(\bx_{\hspa\bm{\alpha}}^{+},\bx_{(\hspa-\hsg)\bm{\alpha}}^{+})
 \hf_{*}(\bx_{(\hspa-\hsg)\bm{\alpha}}^{+})\hf(\bx_{\hspa\bm{\alpha}}^{+}) \\
=&[\hg^{[0]}(\bx,\bx)+\hsg\hg^{[1]}(\bx,\bx)]\hf_*(\bx)\hf(\bx)
 +\frac{\partial\hg^{[0]}(\bx,\by)}{\partial y_i}\big|_{\by=\bx}(2\hspa-\hsg)\alpha_i\hf_*(\bx)\hf(\bx)\notag\\
 & +\hg^{[0]}(\bx,\bx)\hf(\bx)\frac{\partial \hf_*(\bx)}{\partial x_i}(\hspa-\hsg)\alpha_i
 +\hg^{[0]}(\bx,\bx)\hf_*(\bx)\frac{\partial \hf(\bx)}{\partial x_i}\hspa\alpha_i
 +O(\hsg^2) \notag\\
=&[\mG(\bx)+\hsg\hg^{[1]}(\bx,\bx)]\hf_*(\bx)\hf(\bx)
 +\frac12\frac{\partial\mG(\bx)}{\partial x_i}(2\hspa-\hsg)\alpha_i\hf_*(\bx)\hf(\bx)\notag\\
 & +\mG(\bx)\hf(\bx)\frac{\partial \hf_*(\bx)}{\partial x_i}(\hspa-\hsg)\alpha_i
 +\mG(\bx)\hf_*(\bx)\frac{\partial \hf(\bx)}{\partial x_i}\hspa\alpha_i
 +O(\hsg^2).\notag
\end{align}
\noindent
Once this result is substituted into \eqref{eq:pqcol},
the integration with respect to $\hspa$ can be carried out immediately.
Furthermore, using the formulas
\begin{equation}
  \int \alpha_i\alpha_j\hV_\alpha^2\theta(\hV_\alpha)d\Omega(\bm{\alpha})
= \frac{4\pi}{15}(\hV_i\hV_j+\frac12\hbV^2\delta_{ij}),
\end{equation}%
\begin{equation}
  \int \alpha_i\alpha_j\alpha_k\hV_\alpha^2\theta(\hV_\alpha)d\Omega(\bm{\alpha})
= \frac{\pi}{12}
  \Big(
  \frac{\hV_i\hV_j\hV_k}{|\hbV|}+|\hbV|(\hV_i\delta_{jk}+\hV_j\delta_{ki}+\hV_k\delta_{ij})
  \Big),
\end{equation}
\noindent
the integration with respect to $\bm{\alpha}$ can also be carried out to yield
\begin{align}
\hp_{ij}^{\col}= & \frac{\kappa}{2\sqrt{2\pi}}\mG
\int \alpha_{i}\alpha_{j}\hV_{\alpha}^{2}\theta(\hV_{\alpha}) \Big\{ (1+\frac{g^{[1]}}{\mG}\hsg)\hf_*\hf \\
 & +\frac12\hsg(\hf_*\frac{\partial \hf}{\partial x_k}
 -\hf\frac{\partial \hf_*}{\partial x_k})\alpha_k
 \Big\} d\Omega(\bm{\alpha})d\bz_{*}d\bz+O(\hsg^2)
\notag\\
=&\frac{2}{15}\sqrt{\frac{\pi}{2}}\kappa\mG
\int\Big\{
    (\hV_i\hV_j+\frac12\hV^2\delta_{ij})(1+\frac{g^{[1]}}{\mG}\hsg)
    \hf_*\hf 
 \notag\\
 & +\frac{5}{32}\hsg
  \Big(
   \frac{\hV_i\hV_j\hV_k}{\hV}+{\hV}(\hV_i\delta_{jk}+\hV_j\delta_{ki}+\hV_k\delta_{ij})
  \Big)\notag\\
  &\times (\hf_*\frac{\partial \hf}{\partial x_k}
 -\hf\frac{\partial \hf_*}{\partial x_k})
  \Big\} d\bz_{*}d\bz+O(\hsg^2),\notag
\end{align}
\begin{align}
\hq_{i}^{\col}
= & \frac{1}{4\sqrt{2\pi}}\kappa\mG
\int \alpha_{i}[(\hbC+\hbC_{*})\cdot\bm{\alpha}]
     \hV_{\alpha}^{2}\theta(\hV_{\alpha})
  \{(1+\frac{g^{[1]}}{\mG}\hsg)\hf_*\hf \\
 & +\frac12\hsg[\hf_*\frac{\partial \hf}{\partial x_k}
 -\hf\frac{\partial \hf_*}{\partial x_k}]\alpha_k
 \} d\Omega(\bm{\alpha})d\bz_{*}d\bz+O(\hsg^2) \notag \\
= & \frac{1}{15}\sqrt{\frac{\pi}{2}}\kappa\mG
\int (\hC_j+\hC_{*j})\Big\{
    (\hV_i\hV_j+\frac12\hV^2\delta_{ij})(1+\frac{g^{[1]}}{\mG}\hsg)
    \hf_*\hf 
 \notag\\
 & +\frac{5}{32}\hsg
  \Big(
   \frac{\hV_i\hV_j\hV_k}{\hV}+{\hV}(\hV_i\delta_{jk}+\hV_j\delta_{ki}+\hV_k\delta_{ij})
  \Big)\notag\\
  &\times (\hf_*\frac{\partial \hf}{\partial x_k}
 -\hf\frac{\partial \hf_*}{\partial x_k})
  \Big\}  d\bz_{*}d\bz+O(\hsg^2),\notag
\end{align}
\noindent
where the argument $\bx$ has been suppressed in the above expressions.
Substitution of the expansion of $\hf$ [i.e., \eqref{eq:f_exp}] 
and the relation $\hsg=\kappa\ve$ finally yields \eqref{eq:hpq_exps}.

\section{Specific form of $\mG$\label{app:g_specf}}
In this appendix, we briefly discuss the form of $\mG$ and $\hg^{[1]}$. 
Because of the definition \eqref{eq:hg_def},
\[
\hg(\bx,\bx)=2\hmS(\hmR(\bx)),\quad
\frac{\partial\hg(\bx,\by)}{\partial y_i}\Big|_{\by=\bx}
=\frac{\partial\hmS(\hmR(\bx))}{\partial x_i}
=\frac{d\hmS(\hmR(\bx))}{d\hmR(\bx)}
 \frac{\partial\hmR(\bx)}{\partial x_i},
\]
\noindent
in the bulk away from the boundary.$^2$%
\footnote{$^2$ Note that $\chi_\hD(\bx)$ can be regarded as unity if the position $\bx$ is away from the boundary.}
\noindent
However, since 
\[
\hmR(\bx)=\hr(\bx)+O(\hsg^2),
\]
\noindent
by \eqref{eq:hR_def} in the bulk, we have
\begin{align}
& \mG(\bx)\big(=\hg^{[0]}(\bx,\bx)\big)
 \equiv\lim_{\hsg\to0}\hg(\bx,\bx)=2\hmS(\hr(\bx)),
\quad \hg^{[1]}(\bx,\bx)=0,\\
& \frac{\partial\hg^{[0]}(\bx,\by)}{\partial y_i}\Big|_{\by=\bx}
= \frac{\partial\hmS(\hr(\bx))}{\partial x_i}
%= \frac12\frac{\partial\hg^{[0]}(\bx,\bx)}{\partial x_i}
= \frac12\frac{\partial\mG(\bx)}{\partial x_i}.
\end{align}

In the case of the OEE, 
since $\hg(\bx,\by)$ takes the form
\[
\hg(\bx,\by)= \hmY(\hr(\frac{\bx+\by}{2}))
\Big(\equiv\frac{1}{\sfg_0}\mY(\rho_0\hr(\frac{\bx+\by}{2}))\Big),
\]
\noindent
by the definition of \eqref{eq:gY},
the expansion of $\hg$ in terms of $\hsg$ is not necessary, so that
we may put
\begin{align}
& \mG(\bx)\equiv\hg^{[0]}(\bx,\bx)=\hg(\bx,\bx)=\hmY(\hr(\bx)),
\quad \hg^{[1]}(\bx,\bx)=0,\displaybreak[0]\\
& \frac{\partial\hg^{[0]}(\bx,\by)}{\partial y_i}\Big|_{\by=\bx}
= \frac{\partial\hg(\bx,\by)}{\partial y_i}\Big|_{\by=\bx}
= \frac12\frac{\partial\hmY(\hr(\bx))}{\partial x_i},
\end{align}
\noindent
again in the bulk away from the boundary.
Hence, 
as far as the EoS is taken to be common,
i.e., $2\hmS(\hr(\bx))$ and $\hmY(\hr(\bx))$ (or $2\mS(\rho(\bX))$ and $\mY(\rho(\bX))$) are identical  (see Sec.~\ref{sec:Eos}),
$\mG$ is also common between the EESM and the OEE.

\section{Reduction of $\mC_s^{(1)}\partial\hf^{(0)}/\partial\vr_s$ \label{app:Inhom}}

Since $\mC_0^{(1)}=0$, 
the second term on the right-hand side of \eqref{eq:mLhf1} is transformed as
\begin{align}
  \mC_s^{(1)}\frac{\partial\hf^{(0)}}{\partial\vr_s} 
=&\mC_j^{(1)}\frac{2}{\hr\hT}
  \Big[\hC_j-\frac23\hv_j(\frac{\hbC^2}{\hT}-\frac32)\Big]\hf^{(0)}
+ \mC_4^{(1)}\frac23\frac{1}{\hr \hT}
(\frac{\hbC^2}{\hT}-\frac32)\hf^{(0)} \displaybreak[0]\\
=&-\frac{\kappa}{6}\sqrt{\frac{\pi}{2}}
   \frac{\partial(\hr^2\mG\hT)}{\partial x_j}\frac{2}{\hr\hT}
  \Big[\hC_j-\frac23\hv_j(\frac{\hbC^2}{\hT}-\frac32)\Big]\hf^{(0)}\notag\\
&-\frac{\kappa}{3}\sqrt{\frac{\pi}{2}}
   \frac{\partial(\hr^2\mG\hT\hv_j)}{\partial x_j}\frac23\frac{1}{\hr \hT}
(\frac{\hbC^2}{\hT}-\frac32)\hf^{(0)}\notag \displaybreak[0]\\
=&-\frac{\kappa}{3}\sqrt{\frac{\pi}{2}}
   \frac{\partial}{\partial x_j}\Big(\hr^2\mG\hT\Big)\frac{1}{\hr\hT}
  \hC_j\hf^{(0)}\notag\\
&+\frac{\kappa}{3}\sqrt{\frac{\pi}{2}}
   \frac{\partial(\hr^2\mG\hT)}{\partial x_j}\frac{1}{\hr\hT}
  \frac23\hv_j(\frac{\hbC^2}{\hT}-\frac32)\hf^{(0)}\notag\\
&-\frac{\kappa}{3}\sqrt{\frac{\pi}{2}}
   \frac{\partial(\hr^2\mG\hT\hv_j)}{\partial x_j}\frac23\frac{1}{\hr \hT}
(\frac{\hbC^2}{\hT}-\frac32)\hf^{(0)}\notag \displaybreak[0]\\
=&-\frac{\kappa}{3}\sqrt{\frac{\pi}{2}}
   \frac{\partial(\hr\mG)}{\partial x_j}
  \hC_j\hf^{(0)}
  -\frac{\kappa}{3}\sqrt{\frac{\pi}{2}}
  \hr\mG \frac{\partial\ln(\hr\hT)}{\partial x_j}
  \hC_j\hf^{(0)}\notag\\
&-\frac{\kappa}{3}\sqrt{\frac{\pi}{2}}
  \hr\mG
  \frac23\frac{\partial\hv_j}{\partial x_j}(\frac{\hbC^2}{\hT}-\frac32)\hf^{(0)}.
\notag
\end{align}

\section{Reduction of $\mJ^{[1]}(\hf^{(0)},\hf^{(0)})$\label{app:mJ1_hf0_hf0}}

Compared with $C_s^{(1)}\partial\hf^{(0)}/\partial\vr_s$,
the transformation of $\mJ^{[1]}(\hf^{(0)},\hf^{(0)})$
requires more manipulations, which is as follows:
\begin{align}
 &\mJ^{[1]}(\hf^{(0)},\hf^{(0)})  \\
=&\mG
 \int \alpha_i[\frac{\partial \hf_*^{(0)\prime}}{\partial x_i}\hf^{(0)\prime}
              +\frac{\partial \hf_*^{(0)}}{\partial x_i}\hf^{(0)}]
      \frac{\hV_\alpha\theta(\hV_\alpha)}{2\sqrt{2\pi}}d\Omega(\bm{\alpha})d\bz_* \notag\\
&+\frac12\frac{\partial \mG}{\partial x_i}
 \int \alpha_i[\hf_*^{(0)\prime}\hf^{(0)\prime}+\hf_*^{(0)}\hf^{(0)}]
 \frac{\hV_\alpha\theta(\hV_\alpha)}{2\sqrt{2\pi}}d\Omega(\bm{\alpha})d\bz_*
 \notag \\
=&\mG
 \int  2\alpha_i\hf^{(0)}\hf_*^{(0)}
  [\frac{\partial}{\partial x_i} \ln\hr
   +\frac{\hC_{*j}^\prime+\hC_{*j}}{\hT}\frac{\partial\hv_j}{\partial x_i} \notag \\
&+(\frac{\hbC_*^2+\hbC_*^{\prime 2}}{2\hT}-\frac32)\frac{\partial}{\partial x_i}\ln\hT] 
  \frac{\hV_\alpha\theta(\hV_\alpha)}{2\sqrt{2\pi}}
  d\Omega(\bm{\alpha})d\bz_* \notag\\
&+\frac{\partial \mG}{\partial x_i}
  \int \alpha_i \hf_*^{(0)}\hf^{(0)}
  \frac{\hV_\alpha\theta(\hV_\alpha)}{2\sqrt{2\pi}}d\Omega(\bm{\alpha})d\bz_*
   \notag \\
=&\mG
 \int  \alpha_i\hf^{(0)}\hf_*^{(0)}
  [\frac{\partial}{\partial x_i} \ln\hr
   +\frac{2\hC_{*j}}{\hT}\frac{\partial\hv_j}{\partial x_i}
   +(\frac{\hbC_*^2}{\hT}-\frac32)\frac{\partial}{\partial x_i}\ln\hT \notag \\
&  -\frac{\hV_\alpha\alpha_j}{\hT}(\frac{\partial\hv_j}{\partial x_i}
 +\hC_{*j}\frac{\partial}{\partial x_i}\ln\hT)
  +(\frac{\hV_\alpha^2}{2\hT})\frac{\partial}{\partial x_i}\ln\hT] 
  \frac{\hV_\alpha\theta(\hV_\alpha)}{\sqrt{2\pi}}
  d\Omega(\bm{\alpha})d\bz_*  \notag\\
&  -\frac13\sqrt{\frac{\pi}{2}}\hr\hC_i\hf^{(0)}\frac{\partial \mG}{\partial x_i}
   \notag \\
=&\mG
 \int  \hf^{(0)}\hf_*^{(0)}
  [\frac{2\pi}{3}\hV_i\frac{\partial}{\partial x_i} \ln\hr
  +\frac{2\pi}{3}\hV_i\frac{2\hC_{*j}}{\hT}\frac{\partial\hv_j}{\partial x_i} \notag \\
&
  +\frac{2\pi}{3}\hV_i(\frac{\hbC_*^2}{\hT}-\frac32)\frac{\partial}{\partial x_i}\ln\hT \notag \\
&
 -\frac{4\pi}{15}(\hV_i\hV_j+\frac12\hbV^2\delta_{ij})
  \frac{1}{\hT}(\frac{\partial\hv_j}{\partial x_i}
 +\hC_{*j}\frac{\partial}{\partial x_i}\ln\hT)
 \notag\\
&+\frac{2\pi}{5}\hV_i\hbV^2\frac{1}{2\hT}\frac{\partial}{\partial x_i}\ln\hT] 
  \frac{1}{\sqrt{2\pi}} d\bz_* 
  -\frac13\sqrt{\frac{\pi}{2}}\hr\hC_i\hf^{(0)}\frac{\partial \mG}{\partial x_i}
   \notag \\
=&\mG
 \int \hf^{(0)}\hf_*^{(0)}
  [ \frac{4\pi}{3}\hC_{*i}\hC_{*j}\frac{1}{\hT}\frac{\partial\hv_j}{\partial x_i}
%  \notag\\
%&
 -\frac{4\pi}{15}(\hC_{*i}\hC_{*j}+\hC_i\hC_j
 \notag \\
&+\frac12(\hbC_*^2+\hbC^2)\delta_{ij})
  \frac{1}{\hT}\frac{\partial\hv_j}{\partial x_i}
% \notag\\
%&
 +\frac{4\pi}{15}(2\hC_j\hC_{*i}\hC_{*j}+\hC_i\hbC_*^2)
  \frac{1}{\hT}\frac{\partial}{\partial x_i}\ln\hT
 \notag\\
&-\frac{\pi}{5}(\hC_i(\hbC_*^2+\hbC^2)+2\hC_j\hC_{*j}\hC_{*i})
  \frac{1}{\hT}\frac{\partial}{\partial x_i}\ln\hT] 
  \frac{1}{\sqrt{2\pi}} d\bz_* \notag\\
&-\frac13\sqrt{\frac{\pi}{2}}\hC_i\hf^{(0)}
  \Big( 2\mG\frac{\partial\hr}{\partial x_i}
       +\hr\frac{\partial \mG}{\partial x_i}\Big)
   \notag \\
=&\hr\mG\hf^{(0)}
  [ -\frac{4\pi}{15}(\hC_i\hC_j+\frac12(\hbC^2-\frac52\hT)\delta_{ij})
  \frac{1}{\hT}\frac{\partial\hv_j}{\partial x_i}
  +\frac{2\pi}{3}\hC_i\frac{\partial}{\partial x_i}\ln\hT
  \notag\\
&-\frac{\pi}{5}\hC_i(\frac{\hbC^2}{\hT}+\frac52)
  \frac{\partial}{\partial x_i}\ln\hT] 
  \frac{1}{\sqrt{2\pi}} 
  -\frac13\sqrt{\frac{\pi}{2}}\hC_i\hf^{(0)}
  \Big( 2\mG\frac{\partial\hr}{\partial x_i}
       +\hr\frac{\partial \mG}{\partial x_i}\Big)
   \notag \\
=&-\frac13\sqrt{\frac{\pi}{2}} \hr\mG\hf^{(0)}
  \Big[\frac{4}{5}(\frac{\hC_i\hC_j}{\hT}
                  +\frac12(\frac{\hbC^2}{\hT}-\frac52)\delta_{ij})
                  \frac{\partial\hv_j}{\partial x_i}
 \notag\\
&+\frac{3}{5}\hC_i(\frac{\hbC^2}{\hT}-\frac52)
  \frac{\partial}{\partial x_i}\ln\hT 
 +\hC_i\frac{\partial\ln(\hr^2\mG\hT)}{\partial x_i}\Big],\notag
\end{align}
\noindent
where
\begin{align}
& \int \alpha_i\hV_\alpha\theta(\hV_\alpha)d\Omega(\bm{\alpha})
=\frac{2\pi}{3}\hV_i,\\
& \int \alpha_i\hV_\alpha^3\theta(\hV_\alpha)d\Omega(\bm{\alpha})
=\frac{2\pi}{5}\hV_i\hbV^2,
\end{align}
\noindent
have been used.

\section{Derivation of $\hp_{ij}^{\kin(1)}$ and $\hq_i^{\kin(1)}$\label{app:pq_kin1}}

Once $\hf^{(1)}$ is obtained in the form \eqref{eq:hf1_final},
the expressions of $\hp_{ij}^{\kin(1)}$ and $\hq_i^{\kin(1)}$ can be derived essentially by following the procedure
in the case of Boltzmann equation.
Note that the odd and the even part of $\hf^{(1)}$ with respect to $\hbC$ does not contribute to the stress tensor and the heat-flow vector respectively.
Hence, these parts can be omitted from the beginning in the reduction
of the corresponding quantities. Consequently, we have
\begin{align}
 \hp_{ij}^{\kin(1)}
=&2\int \hC_i\hC_j\hf^{(1)}d\bz  \\
=&-\frac{1}{\hr\mG}( 1+\frac25\Lambda )
   \sqrt{\hT} {\frac{\partial \hv_k}{\partial x_l}} \notag\\
 &\times \int\frac{2\hC_i\hC_j}{\hT}
    B(\frac{|\hbC|}{\sqrt{\hT}})
           \frac{1}{\hT} (\hC_k\hC_l-\frac13\hbC^2\delta_{kl})
            \hf^{(0)} d\hbC \notag\\
=&-\frac{1}{\hr\mG}( 1+\frac25\Lambda )
   \hT^2{\frac{\partial \hv_k}{\partial x_l}}
   \frac{8\pi}{15}\int_0^\infty Z^6B(Z)\exp(-Z^2)dZ\notag\\
 & \times\frac{\hr}{(\pi\hT)^{3/2}}(\delta_{ik}\delta_{lj}+\delta_{il}\delta_{jk}-\frac23\delta_{ij}\delta_{kl})\notag\\
=&-\frac{1}{\mG}( 1+\frac25\Lambda )
   \gamma_1\sqrt{\hT}\overline{\frac{\partial \hv_i}{\partial x_j}},\notag
\end{align}
\noindent
and
\begin{align}
 \hq_i^{\kin(1)}
=&\int\hC_i\hbC^2\hf^{(1)}d\bz   \\
=& -\frac{\sqrt{\hT}}{\hr\mG}(1+\frac35\Lambda )
   \frac{\partial\hT}{\partial x_i}
   \int \frac13(\frac{|\hbC|}{\sqrt{\hT}})^4
       A(\frac{|\hbC|}{\sqrt{\hT}})\hf^{(0)}d\hbC \notag\\
=& -\frac{\sqrt{\hT}}{\mG}\frac{4}{3\sqrt{\pi}}(1+\frac35\Lambda )
   \frac{\partial\hT}{\partial x_i}
   \int_0^\infty Z^6 A(Z) \exp(-Z^2) dZ\notag\\
=& -\frac{1}{\mG}(1+\frac35\Lambda )
    \frac{5}{4}\gamma_2\sqrt{\hT}
   \frac{\partial\hT}{\partial x_i}.\notag
\end{align}

\section{Derivation of $\hp_{ij}^{\col(1)}$ and $\hq_i^{\col(1)}$\label{app:pq_col1}}

We first make a reduction of $\hp_{ij}^{\col(1)}$ from \eqref{eq:hp1}
mainly by using that $\langle \psi_s \hf^{(1)}\rangle=0$ and that $\hf^{(0)}$
is even with respect to $\hbC$:
\begin{align}
    \hp_{ij}^{\col(1)}
= & \frac{2}{15}\sqrt{\frac{\pi}{2}}\kappa\mG
    \int \{(\hV_i\hV_j+\frac12\hbV^2\delta_{ij})
          (\hf^{(0)}_*\hf^{(1)}+\hf_*^{(1)}\hf^{(0)}\\
          &+\frac{\kappa\hg^{[1]}}{\mG}\hf_*^{(0)}\hf^{(0)}) \displaybreak[0]       +\frac{5}{32}\kappa
           \Big(\frac{\hV_i\hV_j\hV_k}{|\hbV|}+|\hbV|(\hV_i\delta_{jk}+\hV_j\delta_{ki}\notag\\
           &+\hV_k\delta_{ij})\Big)  (\hf^{(0)}_*\frac{\partial \hf^{(0)}}{\partial x_k}
          -\hf^{(0)}\frac{\partial \hf^{(0)}_*}{\partial x_k})
         \} d\bz_{*}d\bz\notag\displaybreak[0]\\
= & \frac{4}{15}\sqrt{\frac{\pi}{2}}\kappa\mG
    \int (\hC_{*i}\hC_{*j}+\hC_{i}\hC_{j}+\frac12(\hbC_*^2+\hbC^2)\delta_{ij})
           (\hf^{(0)}_*\hf^{(1)}\notag\\
           &\quad+\frac12\frac{\kappa\hg^{[1]}}{\mG}\hf_*^{(0)}\hf^{(0)}) d\bz_{*}d\bz
            +\frac{1}{24}\sqrt{\frac{\pi}{2}}
    \kappa^2\mG
    \int   \Big(\frac{\hV_i\hV_j\hV_k}{|\hbV|}\notag \\
    &\quad+|\hbV|(\hV_i\delta_{jk}+\hV_j\delta_{ki}+\hV_k\delta_{ij})\Big)
            \hf^{(0)}_*\frac{\partial \hf^{(0)}}{\partial x_k}
          d\bz_{*}d\bz\notag\\
\equiv & \hp_{ij}^{\col(1)a}+\hp_{ij}^{\col(1)b}.\notag
\end{align}
\noindent
Then, on one hand, the $a$-part is readily transformed as
\begin{align}
    \hp_{ij}^{\col(1)a}
= & \frac{4}{15}\sqrt{\frac{\pi}{2}}\kappa\mG
    \int (\hC_{*i}\hC_{*j}+\hC_{i}\hC_{j}+\frac12(\hbC_*^2+\hbC^2)\delta_{ij})
          (\hf^{(0)}_*\hf^{(1)}  \\
          &+\frac12\frac{\kappa\hg^{[1]}}{\mG}\hf_*^{(0)}\hf^{(0)}) d\bz_{*}d\bz \notag\displaybreak[0]\\
= & \frac{4}{15}\sqrt{\frac{\pi}{2}}\kappa\hr\mG
    \int ( \hC_{i}\hC_{j}+\frac12\hbC^2\delta_{ij}+\frac54\hT\delta_{ij})
    (\hf^{(1)}+\frac12\frac{\kappa\hg^{[1]}}{\mG}\hf^{(0)})d\bz \notag\displaybreak[0]\\
= & \frac{4}{15}\sqrt{\frac{\pi}{2}}\kappa\hr
    \Big(\mG\int \hC_{i}\hC_{j}\hf^{(1)}d\bz 
    +\frac{5}{4}\kappa\hg^{[1]}\hr\hT\delta_{ij}
    \Big)\notag\displaybreak[0]\\
= & \frac{2}{15}\sqrt{\frac{\pi}{2}}\kappa\hr
    \Big(\mG \hp_{ij}^{\mathrm{kin}(1)}
    +\frac{5}{2}\kappa\hg^{[1]}\hr\hT\delta_{ij}
    \Big).\notag
\end{align}
\noindent
On the other hand, the $b$-part
needs a change of variables from $(\bz,\bz_*)$ to $(\hbV,\hbU)$
with $\hbV=\hbC_*-\hbC$ and $\hbU=\hbC_*+\hbC$ for further reductions.
Since $d\bz d\bz_*=(1/8)d\hbU d\hbV$, 
it holds that
\begin{align}        
    \hp_{ij}^{\col(1)b}
  =&\frac{1}{24}\sqrt{\frac{\pi}{2}}
    \kappa^2\mG
    \int   \Big(\frac{\hV_i\hV_j\hV_k}{|\hbV|} \\
    &\qquad +|\hbV|(\hV_i\delta_{jk}+\hV_j\delta_{ki}+\hV_k\delta_{ij})\Big)
            \hf^{(0)}_*\frac{\partial \hf^{(0)}}{\partial x_k}
          d\bz_{*}d\bz \notag\displaybreak[0]\\
  =&\frac{1}{192}\sqrt{\frac{\pi}{2}}
    \kappa^2\mG
    \int  \Big(\frac{\hV_i\hV_j\hV_k}{|\hbV|}+|\hbV|(\hV_i\delta_{jk}+\hV_j\delta_{ki}+\hV_k\delta_{ij})\Big) \notag \\
 &          \times\frac{\hr^2}{(\pi\hT)^3} \Big(
               \frac{\partial\ln\hr}{\partial x_k} 
              +\big(
               \frac{(\hbU-\hbV)^2}{4\hT}-\frac32
               \big)
               \frac{\partial\ln\hT}{\partial x_k}\notag\\
 &\qquad
              +\frac{\hU_l-\hV_l}{\hT}\frac{\partial\hv_l}{\partial x_k}
            \Big)\exp(-\frac{\hbV^2+\hbU^2}{2\hT})
          d\hbV d\hbU \notag\displaybreak[0]\\
  =&-\frac{1}{192}\sqrt{\frac{\pi}{2}}
    \kappa^2\mG
    \int  \Big(\frac{\hV_i\hV_j\hV_k}{|\hbV|}+|\hbV|(\hV_i\delta_{jk}+\hV_j\delta_{ki}+\hV_k\delta_{ij})\Big)  \notag \\
 &          \times \frac{\hr^2}{(\pi\hT)^3} 
             \frac{\hV_l}{\hT}\frac{\partial\hv_l}{\partial x_k}
             \exp(-\frac{\hbV^2+\hbU^2}{2\hT})
          d\hbV d\hbU \notag\displaybreak[0]\\
  =&-\frac{\sqrt{\hT}}{12\pi}\kappa^2\hr^2\mG
     \frac{\partial\hv_l}{\partial x_k}
    (\frac{4\pi}{15}+\frac{4\pi}{3})(\delta_{ij}\delta_{kl}+\delta_{ik}\delta_{jl}+\delta_{il}\delta_{jk}) \notag\displaybreak[0]\\
  =&-\frac{2}{15}\sqrt{\hT}\kappa^2\hr^2\mG
     (\overline{\frac{\partial\hv_i}{\partial x_j}}
     +\frac53\frac{\partial\hv_k}{\partial x_k}\delta_{ij}).\notag
\end{align}
\noindent
Consequently, since $\hg^{[1]}=0$ for both EESM and OEE (see Appendix~\ref{app:g_specf}), we have
\begin{align}        
    \hp_{ij}^{\col(1)}
& = \frac{2}{15}\sqrt{\frac{\pi}{2}}\kappa\hr\mG
    \hp_{ij}^{\kin(1)} 
    -\frac{2}{15}\sqrt{\hT}\kappa^2\hr^2\mG
     (\overline{\frac{\partial\hv_i}{\partial x_j}}
     +\frac53\frac{\partial\hv_k}{\partial x_k}\delta_{ij})\\
& =  \frac{2}{5}\Lambda \hp_{ij}^{\kin(1)} 
    -\frac{12}{5\pi}\frac{\sqrt{\hT}}{\mG}\Lambda^2
     \overline{\frac{\partial\hv_i}{\partial x_j}}
    -\frac{4}{\pi}\frac{\sqrt{\hT}}{\mG}\Lambda^2
     \frac{\partial\hv_k}{\partial x_k}\delta_{ij},\notag
\end{align}
\noindent
which is identical to \eqref{eq:hp_col1_final}.

In the same way, we first transform $\hq_i^{\col(1)}$ from \eqref{eq:hq1} as
\begin{align}
\hq_{i}^{\col(1)}
= & \frac{1}{15}\sqrt{\frac{\pi}{2}}\kappa\mG
\int (\hC_j+\hC_{*j})\Big\{
    (\hV_i\hV_j+\frac12\hbV^2\delta_{ij})
    (\hf_*^{(1)}\hf^{(0)}+\hf_*^{(0)}\hf^{(1)}\\
    & +\frac{\kappa\hg^{[1]}}{\mG}\hf^{(0)}\hf_*^{(0)})
     +\frac{5}{32}\kappa
  \Big(
   \frac{\hV_i\hV_j\hV_k}{|\hbV|}+|\hbV|(\hV_i\delta_{jk}+\hV_j\delta_{ki}
   \notag \\
   &+\hV_k\delta_{ij})
  \Big) (\hf_*^{(0)}\frac{\partial \hf^{(0)}}{\partial x_k}
 -\hf^{(0)}\frac{\partial \hf_*^{(0)}}{\partial x_k})
  \Big\} d\bz_{*}d\bz \notag \displaybreak[0]\\
= & \frac{2}{15}\sqrt{\frac{\pi}{2}}\kappa\mG
\int (\hC_j+\hC_{*j})
    (\hV_i\hV_j+\frac{\hbV^2}{2}\delta_{ij})
    (\hf_*^{(1)}\hf^{(0)}\notag\\
    &+\frac{\kappa\hg^{[1]}}{2\mG}\hf^{(0)}\hf_*^{(0)}) d\bz_{*}d\bz
     -\frac{1}{48}\sqrt{\frac{\pi}{2}}\kappa^2\mG
\int
  \Big(
   \frac{\hV_i\hV_j\hV_k}{|\hbV|}\notag\\
   &+|\hbV|(\hV_i\delta_{jk}+\hV_j\delta_{ki}+\hV_k\delta_{ij})
  \Big) (\hC_j+\hC_{*j}) \hf^{(0)}\frac{\partial \hf_*^{(0)}}{\partial x_k}
   d\bz_{*}d\bz \notag \\
= & \frac{2}{15}\sqrt{\frac{\pi}{2}}\kappa\mG
\int (\hC_j+\hC_{*j})
    (\hV_i\hV_j+\frac{\hbV^2}{2}\delta_{ij})
    \hf_*^{(1)}\hf^{(0)} d\bz_{*}d\bz\notag\\
    &-\frac{1}{48}\sqrt{\frac{\pi}{2}}\kappa^2\mG
\int
  \Big(
   \frac{\hV_i\hV_j\hV_k}{|\hbV|}+|\hbV|(\hV_i\delta_{jk}+\hV_j\delta_{ki}
   \notag\\
   &+\hV_k\delta_{ij})
  \Big) (\hC_j+\hC_{*j}) \hf^{(0)}\frac{\partial \hf_*^{(0)}}{\partial x_k}
   d\bz_{*}d\bz \notag \\
\equiv & \hq_i^{\col(1)a}+\hq_i^{\col(1)b},\notag
\end{align}
\noindent
where the contribution from $\hg^{[1]}$ has already vanished,
irrespective of its value.
The $a$-part is further transformed as
\begin{align}
\hq_i^{\col(1)a}
= &\frac{2}{15}\sqrt{\frac{\pi}{2}}\kappa\mG
   \int (\hC_j+\hC_{*j})
    (\hV_i\hV_j+\frac12\hV^2\delta_{ij})
    \hf^{(1)}_*\hf^{(0)} d\bz_{*}d\bz  \displaybreak[0]\\
= &\frac{2}{15}\sqrt{\frac{\pi}{2}}\kappa\mG
   \int (\hC_j+\hC_{*j})
    [(\hC_{*i}-\hC_i)(\hC_{*j}-\hC_j)\notag \\
    & \quad +\frac12(\hbC_*-\hbC)^2\delta_{ij}]
    \hf^{(1)}_*\hf^{(0)} d\bz_{*}d\bz \notag\displaybreak[0]\\
= &\frac{2}{15}\sqrt{\frac{\pi}{2}}\kappa\mG
   \int 
    [(\hC_{*i}-\hC_i)(\hbC_*^2-\hbC^2)\notag \\
    &\quad +\frac12(\hC_i+\hC_{*i})(\hbC_*-\hbC)^2]
    \hf^{(1)}_*\hf^{(0)} d\bz_{*}d\bz \notag\displaybreak[0]\\
= &\frac{2}{15}\sqrt{\frac{\pi}{2}}\kappa\mG
   \int 
    [\hC_{*i}(\hbC_*^2-\hbC^2)\notag \\
    &\quad +\frac12(\hC_i+\hC_{*i})(\hbC_*^2+\hbC^2-2\hC_{*l}\hC_l)]
    \hf^{(1)}_*\hf^{(0)} d\bz_{*}d\bz \notag\displaybreak[0]\\
= &\frac{2}{15}\sqrt{\frac{\pi}{2}}\kappa\mG
   \int 
    [\hC_{*i}(\hbC_*^2-\hbC^2)+\frac12\hC_i(-2\hC_{*l}\hC_l)\notag \\
    &\quad +\frac12\hC_{*i}(\hbC_*^2+\hbC^2)]
    \hf^{(1)}_*\hf^{(0)} d\bz_{*}d\bz \notag\displaybreak[0]\\
= &\frac{2}{15}\sqrt{\frac{\pi}{2}}\kappa\mG
   \int 
    [\hC_{*i}\hbC_*^2+\frac12\hC_{*i}\hbC_*^2]
    \hf^{(1)}_*\hf^{(0)} d\bz_{*}d\bz \notag\displaybreak[0]\\
= &\frac{1}{5}\sqrt{\frac{\pi}{2}}\kappa\mG
   \int \hC_{*i}\hbC_*^2
    \hf^{(1)}_*\hf^{(0)} d\bz_{*}d\bz \notag\displaybreak[0]\\
= &\frac{1}{5}\sqrt{\frac{\pi}{2}}\kappa\hr\mG\hq_i^{\kin(1)}
   =\frac{3}{5}\Lambda \hq_i^{\kin(1)},\notag
\end{align}
\noindent
while the $b$-part is, again by the change of variables, further reduced to
\begin{align}
\hq_i^{\col(1)b}
=&-\frac{1}{48}\sqrt{\frac{\pi}{2}}\kappa^2\mG
\int
  \Big(
   \frac{\hV_i\hV_j\hV_k}{|\hbV|}+|\hbV|(\hV_i\delta_{jk}+\hV_j\delta_{ki}+\hV_k\delta_{ij})
  \Big) \\
  &\quad \times  \hU_j\hf^{(0)}\frac{\partial \hf_*^{(0)}}{\partial x_k}
   d\bz_{*}d\bz \notag \displaybreak[0]\\
= & -\frac{1}{48}\sqrt{\frac{\pi}{2}}\kappa^2\mG
\int 
 (\frac{\hV_i\hV_j\hV_k}{|\hbV|}+|\hbV|(\hV_i\delta_{jk}+\hV_j\delta_{ik}+\hV_k\delta_{ij}))\notag\\
&\quad\times \hU_j
 \frac{\hr^2}{\pi^3}\exp(-\frac{\hbV^2+\hbU^2}{2\hT}) \big(
  \frac{\hU_l}{\hT}\frac{\partial \hv_l}{\partial x_k}
 +\frac{\hV_l\hU_l}{2\hT}\frac{\partial \ln\hT}{\partial x_k}\big)
  \frac{d\hbU d\hbV}{8\hT^3} \notag\displaybreak[0]\\
= & -\frac{1}{48}\sqrt{\frac{\pi}{2}}\kappa^2\mG
\int \hU_j
 \frac{\hr^2}{\pi^3}\exp(-\frac{\hbV^2+\hbU^2}{2\hT})
 (\frac{\hV_i\hV_j\hV_k}{|\hbV|} \notag\\
&\quad
 +|\hbV|(\hV_i\delta_{jk}+\hV_j\delta_{ik}+\hV_k\delta_{ij}))
  \frac{\hV_l\hU_l}{2\hT}\frac{\partial \ln\hT}{\partial x_k}
  \frac{d\hbU d\hbV}{8\hT^3} \notag\displaybreak[0]\\
= & -\frac{\sqrt{\pi}}{48}\kappa^2\mG
\int 
 \frac{\hr^2}{\pi^{3/2}}\exp(-\frac{\hbV^2}{2\hT}) \notag\\
&\times 
 (\frac{\hV_i\hV_j\hV_k}{|\hbV|}+|\hbV|(\hV_i\delta_{jk}+\hV_j\delta_{ik}+\hV_k\delta_{ij}))
  \hV_j\frac{\partial \ln\hT}{\partial x_k}
  \frac{d\hbV}{8\hT^{3/2}} \notag\displaybreak[0]\\
= & -\frac{1}{48\pi}\kappa^2\mG
  \int \hr^2\exp(-\frac{\hbV^2}{2\hT}) 
  |\hbV|(3\hV_i\hV_k+\hbV^2\delta_{ik})
  \frac{\partial \ln\hT}{\partial x_k}
  \frac{d\hbV}{8\hT^{3/2}} \notag\displaybreak[0]\\
= & -\frac{1}{6}\kappa^2\hr^2\mG\hT^{3/2}
  \frac{\partial \ln\hT}{\partial x_i} 
= -\frac{3}{\pi}\frac{\Lambda^2}{\mG}\sqrt{\hT}
  \frac{\partial \hT}{\partial x_i}.\notag
\end{align}
\noindent
Consequently, we have
\begin{equation}
\hq_i^{\col(1)}
=\frac{3}{5}\Lambda \hq_i^{\kin(1)}
-\frac{3}{\pi}\frac{\Lambda^2}{\mG}\sqrt{\hT}
  \frac{\partial \hT}{\partial x_i},
\end{equation}
\noindent
which is identical to \eqref{eq:hq_col1final}.

%Set below text at page 2
\lhead[\small\thepage\fancyplain{}\leftmark\hfil$[$]{}
\rhead[]{\small \fancyplain{}\rightmark\small\thepage} \cfoot{}
\markboth {\hfill{\small \rm \Bauthor}\hfill} {\hfill {\small \rm
\Bshorttitle} \hfill}

\renewcommand\bibname{\centerline{\large\bf References}}
\fontsize{9}{11.0pt plus1pt minus .8pt}\selectfont


\begin{thebibliography}{99}
\markboth {\hfill{\small \rm \Bauthor}\hfill} {\hfill {\small \rm
\Bshorttitle} \hfill}

%
%\bibitem{fn1}
%To be precise, it is necessary to make the argument of the logarithmic function dimensionless, like $\ln (f/c_0)$ with a constant $c_0$ having the same dimension as $f$.
%We, however, leave the argument dimensional 
%to avoid additional calculations that do not affect the results.
%
\bibitem{BLPT91}
\newblock N. Bellomo, M. Lachowicz, J. Polewczak and G. Toscani,
\newblock \href{https://doi.org/10.1142/1209}{\emph{Mathematical Topics in Nonlinear Kinetic Theory II}},
\newblock World Scientific, Singapore, 1991. 
%
\bibitem{BB18}
\newblock E. S. Benilov and M. S. Benilov, 
\newblock \href{https://doi.org/10.1103/PhysRevE.97.062115}{Energy conservation and H theorem for the Enskog--Vlasov equation}, 
\newblock \emph{Phys. Rev. E} \textbf{97}, 062115 (2018). 
%
\bibitem{BB19} 
\newblock E. S. Benilov and M. S. Benilov,
\newblock \href{https://doi.org/10.1088/1742-5468/ab3ccf}{The Enskog--Vlasov equation: a kinetic model describing gas, liquid, and solid},
\newblock \emph{J. Stat. Mech.} \textbf{2019}, 103205 (2019).
%
\bibitem{CS69}
\newblock N. F. Carnahan and K. E. Starling,
\newblock \href{https://doi.org/10.1063/1.1672048}{Equation of state for non-attracting rigid spheres},
\newblock \emph{J. Chem. Phys.} \textbf{51}, 635--636 (1969).
%
%\bibitem{CL71}
%\newblock C. Cercignani and M. Lampis, 
%\newblock \href{https://doi.org/10.1080/00411457108231440}{Kinetic models for gas--surface interactions}, 
%\newblock \emph{Trans. Theory Stat. Phys.} \textbf{1}, 101--114  (1971). 
%
%\bibitem{C88}
%\newblock C. Cercignani, 
%\newblock \emph{The Boltzmann Equation and Its Applications}, Springer, New York, 1988.
%
\bibitem{CL88}
\newblock C. Cercignani and M. Lampis, 
\newblock \href{https://doi.org/10.1007/BF01014218}{On the kinetic theory of a dense gas of rough spheres},
\newblock \emph{J. Stat. Phys.} \textbf{53}, 655--672 (1988).
%
%\bibitem{CIP94}
%\newblock C. Cercignani, R. Illner, and M. Pulvirenti,
%\newblock \href{https://doi.org/10.1007/978-1-4419-8524-8}{\emph{The Mathematical Theory of Dilute Gases}}, Springer, New York, 1994, Sec.~9.4.
%
%\bibitem{DG66}
%\newblock J. S. Darrozes and J. P. Guiraud, 
%\newblock \href{https://gallica.bnf.fr/ark:/12148/bpt6k6238594s/f400.image.r=Darrouzes?rk=21459;2}{G\'{e}n\'{e}ralisation formelle du th\'{e}or\`{e}me H en pr\'{e}sence de parois. Applications},
%\newblock \emph{C. R. Acad. Sci. Paris A} \textbf{262}, 1368--1371 (1966).
%
\bibitem{DVK21}
\newblock J. R. Dorfman, H. van Beijeren, and T. R. Kirkpatrick,
\newblock \href{https://doi.org/10.1017/9781139025942}{\emph{Contemporary Kinetic Theory of Matter}}, 
\newblock Cambridge University Press, Cambridge, 2021.
%
\bibitem{E72}
\newblock D. Enskog, 
\newblock \href{https://doi.org/10.1016/C2013-0-02467-X}{Kinetic theory of heat conduction, viscosity,
and self-diffusion in compressed gases and liquids}, 
\newblock in \emph{Kinetic Theory}, Vol. 3, S. G. Brush ed., Pergamon Press, Oxford, Part 2, 1972, pp.226--259.
%
%\bibitem{F97} 
%\newblock A. Frezzotti, 
%\newblock \href{https://doi.org/10.1063/1.869247}{A particle scheme for the numerical solution of the Enskog equation}, 
%\newblock \emph{Phys. Fluids} \textbf{9}, 1329--1335 (1997). 
%
\bibitem{F99}
\newblock A. Frezzotti, 
\newblock \href{https://doi.org/10.1016/S0997-7546(99)80008-9}{Monte Carlo simulation of the heat flow in a dense hard sphere gas}, \newblock \emph{Eur. J. Mech. B/Fluids} \textbf{18}, 103--119 (1999).
%
\bibitem{FGLS18}
\newblock A. Frezzotti, L. Gibelli, D. A. Lockerby, and J. E. Sprittles,
\newblock \href{https://link.aps.org/doi/10.1103/PhysRevFluids.3.054001}{Mean-field kinetic theory approach to evaporation of a binary liquid into vacuum},
\newblock \emph{Phys. Rev. Fluids} \textbf{3}, 054001 (2018). 
%
\bibitem{G58}
\newblock H. Grad,
\newblock \href{https://doi.org/10.1007/978-3-642-45892-7_3}{Principles of the kinetic theory of gases},
\newblock in \emph{Handbuch der Physik}, S. Fl\"{u}gge ed., 
\newblock \textbf{XII}, Springer--Verlag, Berlin, 1958, pp.~205--294.
%
\bibitem{G71}
\newblock M. Grmela, 
\newblock \href{https://doi.org/10.1007/BF01011389}{Kinetic equation approach to phase transitions},
\newblock \emph{J. Stat. Phys.} \textbf{3}, 347--364 (1971).
%
%\bibitem{HTT22}
%\newblock M. Hattori, S. Tanaka and S. Takata,
%\newblock  \href{https://doi.org/10.1063/5.0091390}{Heat transfer in a dense gas between two parallel plates}, 
%\newblock \emph{AIP Advances} \textbf{12}, 055323 (2022).
%
%\bibitem{KKW14}
%\newblock M. Kon, K. Kobayashi, and M. Watanabe, 
%\newblock \href{https://doi.org/10.1063/1.4890523}{Method of determining kinetic boundary conditions in net evaporation/condensation}, 
%\newblock \emph{Phys. Fluids} \textbf{26}, 072003 (2014). 
%
\bibitem{MGB18}
\newblock P. Maynar, M. I. Garcia de Soria, and J. J. Brey,
\newblock \href{https://doi.org/10.1007/s10955-018-1971-7}{The Enskog equation for confined elastic hard spheres}, 
\newblock \emph{J. Stat. Phys.} \textbf{170}, 999--1018 (2018). 
%
\bibitem{R78}
\newblock P. Resibois, 
\newblock \href{https://doi.org/10.1007/BF01011771}{H-theorem for the (modified) nonlinear Enskog equation},
\newblock \emph{J. Stat. Phys.} \textbf{19}, 593--609 (1978).
%
\bibitem{S07}
\newblock Y. Sone, 
\newblock \href{https://doi.org/10.1007/978-0-8176-4573-1}{\emph{Molecular Gas Dynamics}}, Birkh\"{a}user, Boston, 2007.
\newblock  Supplement is available from \href{http://hdl.handle.net/2433/66098}{http://hdl.handle.net/2433/66098}.
%
\bibitem{T24}
\newblock S. Takata, 
\newblock \href{https://doi.org/10.3934/krm.2023025}{On the thermal relaxation of a dense gas described by the modified Enskog equation in a closed system in contact with a heat bath},
\newblock \emph{Kinetic \& Related Models} \textbf{17}, 331--346 (2024).
%
%\bibitem{TT24}
%\newblock S. Takata and A. Takahashi,
%\newblock Note on the summational invariant and corresponding local Maxwellian for the Enskog equation,
%\newblock Kinetic \& Related Models \textbf{17}, 739--754 (2024).
%
\bibitem{TT25}
\newblock S. Takata and A. Takahashi,
\newblock \href{https://doi.org/10.1103/3jnf-mw6y}{Enskog and Enskog-Vlasov equations with a modified correlation factor and their H theorem},
\newblock \emph{Phys. Rev. E} \textbf{111}, 065108 (2025). (\href{https://doi.org/10.48550/arXiv.2502.04744}{arXiv:2502.04744})
%
\bibitem{BE73} 
\newblock H. van Beijeren and M. H. Ernst, 
\newblock \href{https://doi.org/10.1016/0031-8914(73)90372-8}{The modified Enskog equation}, 
\newblock \emph{Physica} \textbf{68}, 437--456 (1973). 
%
\bibitem{vdW}
\newblock J. D. van der Waals,
\newblock Over de Continu\"{i}teit van den Gas -- en Vloeistoftoestand,
Academisch Proefschrift, Leiden (1873) (in Dutch);
\newblock see also the English translation: 
\newblock R. Threlfall and J.F. Adair,
\newblock \emph{Physical Memoirs} \textbf{1}, 333--496 (1890).
%
%\bibitem{WLRZ16} 
%\newblock L. Wu, H. Liu, J. M. Reese, and Y. Zhang,
%\newblock \href{https://doi.org/10.1017/jfm.2016.173}{Non-equilibrium dynamics of dense gas under tight confinement}, 
%\newblock \emph{J. Fluid Mech.} \textbf{794}, 252--266 (2016).
% 
\end{thebibliography}
\end{document}